\documentclass[twocolumn,showpacs,preprintnumbers,amsmath,amssymb,pre,nofootinbib]{revtex4}

\usepackage{graphics}
\usepackage{subfigure}
\newcommand{\partxy}[2]{\ensuremath{\frac{\partial {#1}}{\partial {#2}}}}
\newcommand{\doublepict}[7]
{
 \begin{figure}[htb]\centering
    \subfigure[ ] {
       \resizebox{#7\textwidth}{!}{\includegraphics{#1}}
            \label{fg:#2}
    }
    \subfigure[ ]{
           \resizebox{#7\textwidth}{!}{\includegraphics{#3}}
        \label{fg:#4}
    }
    \caption{#5 \label{fg:#6}}
 \end{figure}
}

\begin{document}
\title{Thermodynamic constraints on fluctuation phenomena}
\author{O J E Maroney}\email{o.maroney@usyd.edu.au}
\affiliation{The Centre for Time and The School of Physics \\
University of Sydney NSW 2006 Australia}
\affiliation{Perimeter Institute for Theoretical Physics \\
31 Caroline St N, Waterloo, ON, N2L 2Y5, Canada}
\date{\today}

\begin{abstract}
The relationships between reversible Carnot cycles, the absence of perpetual motion machines and the existence of a non-decreasing, globally unique entropy function forms the starting point of many textbook presentations of the foundations of thermodynamics.  However, the thermal fluctuation phenomena associated with statistical mechanics has been argued to restrict the domain of validity of this basis of the second law of thermodynamics.  Here we demonstrate that fluctuation phenomena can be incorporated into the traditional presentation, extending, rather than restricting, the domain of validity of the phenomenologically motivated second law.  Consistency conditions lead to constraints upon the possible spectrum of thermal fluctuations. In a special case this uniquely selects the Gibbs canonical distribution and more generally incorporates the Tsallis distributions.  No particular model of microscopic dynamics need be assumed.
\end{abstract}
\maketitle
\section{Introduction}\label{s:intro}
The existence of a globally unique entropy as a function of thermodynamic state, which is non-decreasing in time, is one of the central tenets of classical phenomenological thermodynamics\cite{LY1999,Uff01}.  By contrast, the meaning of entropy within the context of statistical mechanics seems to defy consensus(see \cite{Sheehan2000,BGH2008} for examples).  Since the start of statistical mechanics there has been concern that the existence of fluctuation phenomena leads to violations of the second law of thermodynamics.  This may lead to decreases in entropy, the existence of perpetual motion machines or maybe even the inability to define an entropy at all.  Maxwell's demon represents a persistent strand of thought experiments dedicated to exploring these possibilities\cite{EN98,EN99,LR03}.

Most attempts to construct a second law of thermodynamics for statistical mechanics involve one of two strategies: restrict the domain of validity of the classical statement (usually to reliable, continuous processes) so as to exclude fluctuation phenomena; or to attempt to derive a new second law within the domain of statistical mechanics.  Here we investigate the possibility of a third approach: to extend the domain of the phenomenological second law to include, constrain, and predict the extent of the fluctuation phenomena, which reduces to the more familiar version if fluctuation phenomena are absent.  We find that such an extension seems, in principle, possible, and that with additional work it is possible to define an entropy function consistent with this.  Some possible relationships of this fluctuation second law to conventional statistical mechanics can be inferred.

The approach of the paper is as follows.  Section 2 briefly reviews the equivalence of the Kelvin, Clausius and Carnot versions of the second law of thermodynamics.  Section 3 then proposes an extension of the Kelvin version, to incorporate fluctuation phenomena.  Logically equivalent generalisations of the Clausius and Carnot versions are deduced, and some constraints are deduced about the form of the extended second law.  Section 4 reviews the derivation of an entropy function and shows when the existence of a fluctuation entropy function can be deduced.  Finally Section 5 considers some relationships to statistical mechanical entropies, including the Gibbs and Tsallis\cite{Tsallis1988} entropies.
\section{Phenomenological Second Law}\label{s:phenom}
\begin{figure}[htb]\centering
    \subfigure[ ] {
       \resizebox{0.2\textwidth}{!}{\includegraphics{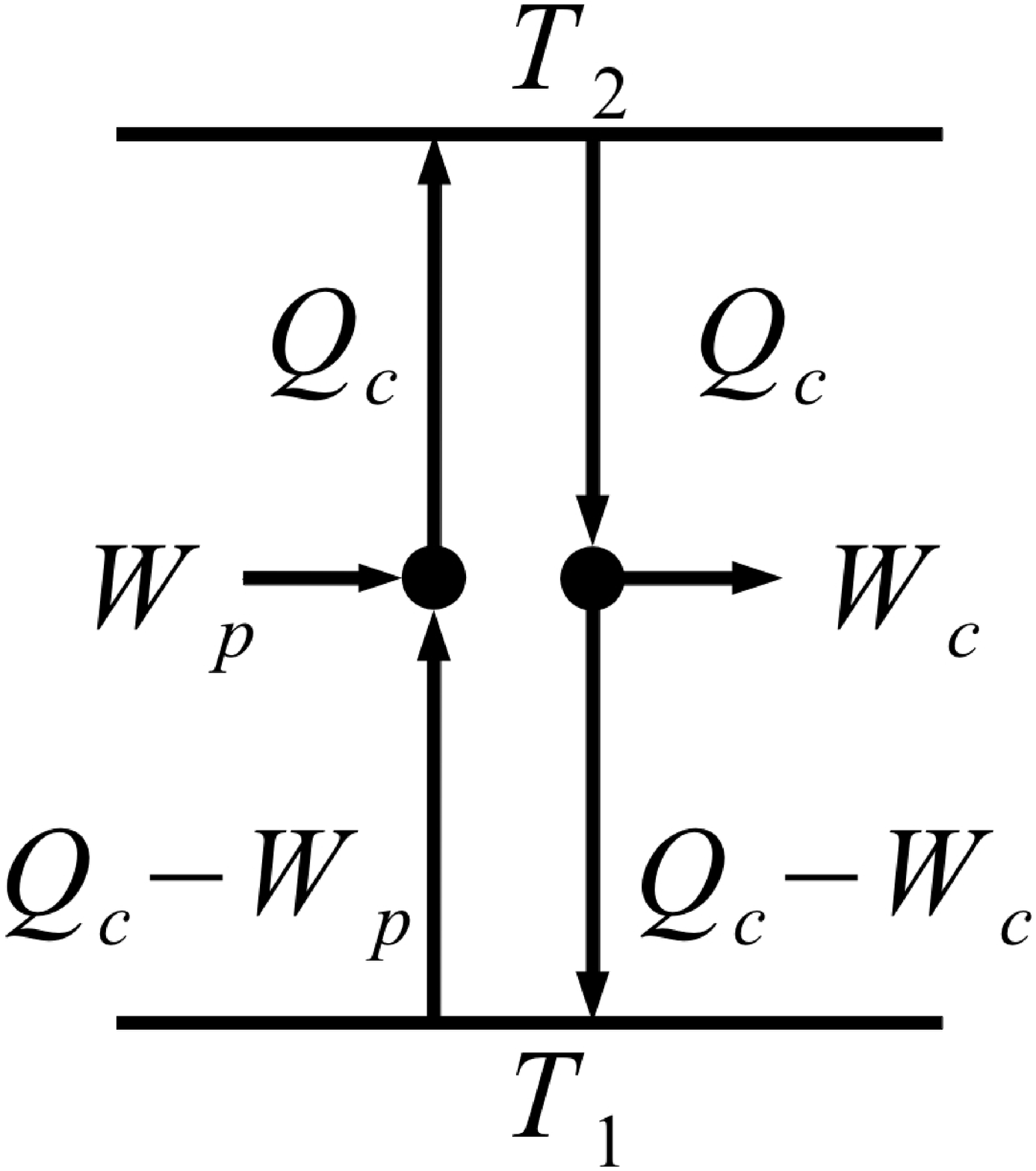}}
            \label{fg:vpumpkelv}
    }
    \subfigure[ ]{
           \resizebox{0.2\textwidth}{!}{\includegraphics{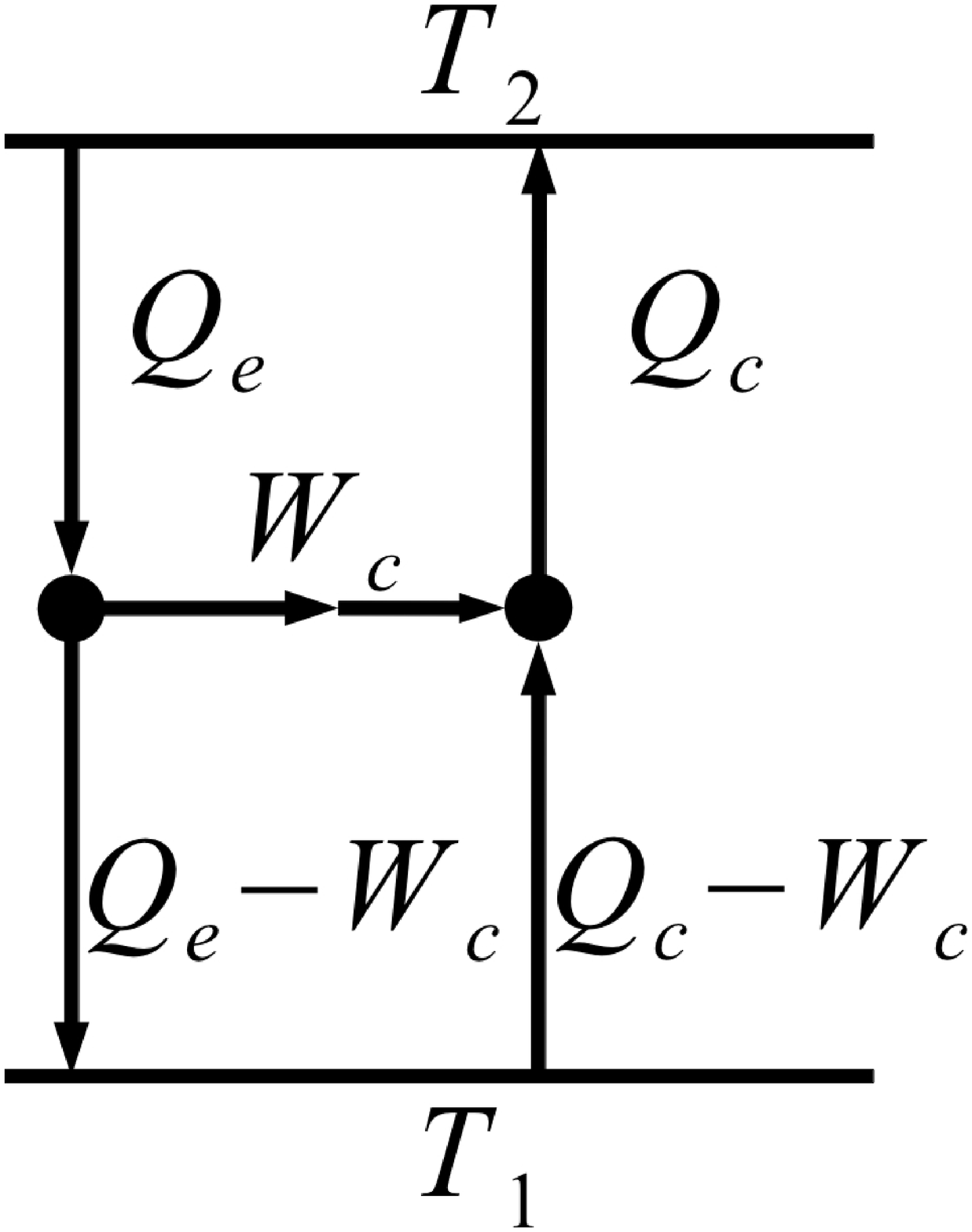}}
        \label{fg:vengclaus}
    }
    \caption{Equivalencies of Violations of Second Laws \label{fg:carnotequiv}}
\end{figure}

Textbook versions of the Second Law of Thermodynamics (see, for example, \cite{Fer1937,Adk68}), when expressed in terms of heat flows and heat baths, take forms such as:
\begin{itemize}
\item \textbf{Kelvin}:  No process is possible whose sole result is the extraction of heat from a heat bath and its conversion to work.
\item \textbf{Clausius}:  No process is possible whose sole result is the transfer of heat from one heat bath to another heat bath at a higher temperature.
\item \textbf{Carnot Heat Engine}:  No heat engine operating between heat baths at temperatures $T_1<T_2$ can operate at an efficiency $n_E$ exceeding the efficiency of a reversible heat engine: $n_E \leq n_{CE}=1- {T_1 \over T_2}$
\item \textbf{Carnot Heat Pump}:  No heat pump operating between heat baths at temperatures $T_1<T_2$ can operate at an efficiency $n_P$ exceeding the efficiency of a reversible heat pump: $n_P \leq n_{CP}={T_2 \over {T_2-T_1}}$
\end{itemize}
 \begin{figure}[htb]\centering
    \subfigure[ $n_{P}={Q_p \over W_p} \leq n_{CP}$ ]{
           \resizebox{0.2\textwidth}{!}{\includegraphics{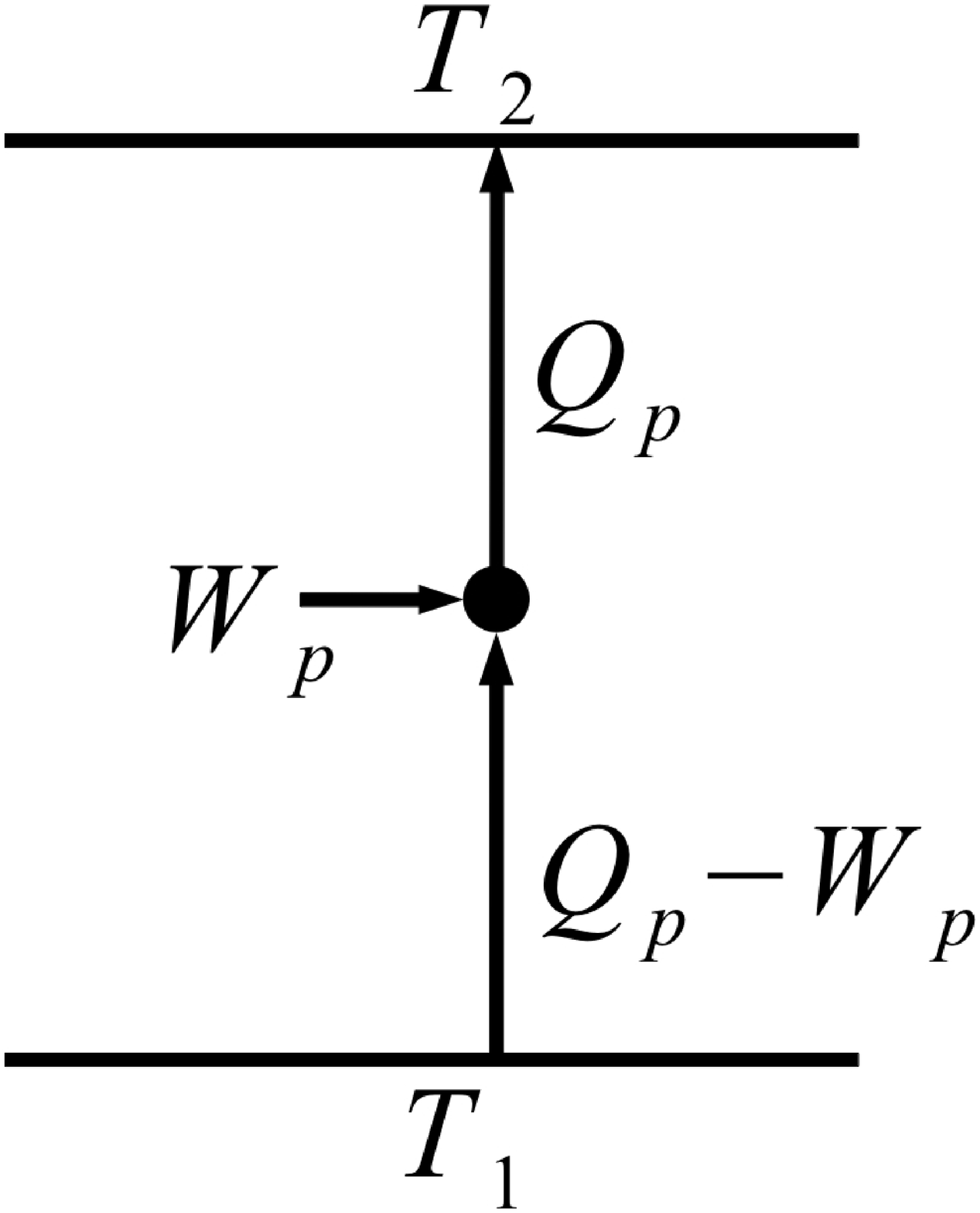}}
        \label{fg:carnotpump}
    }
    \subfigure[ $n_{E}={W_e \over Q_e} \leq n_{CE}$ ] {
       \resizebox{0.2\textwidth}{!}{\includegraphics{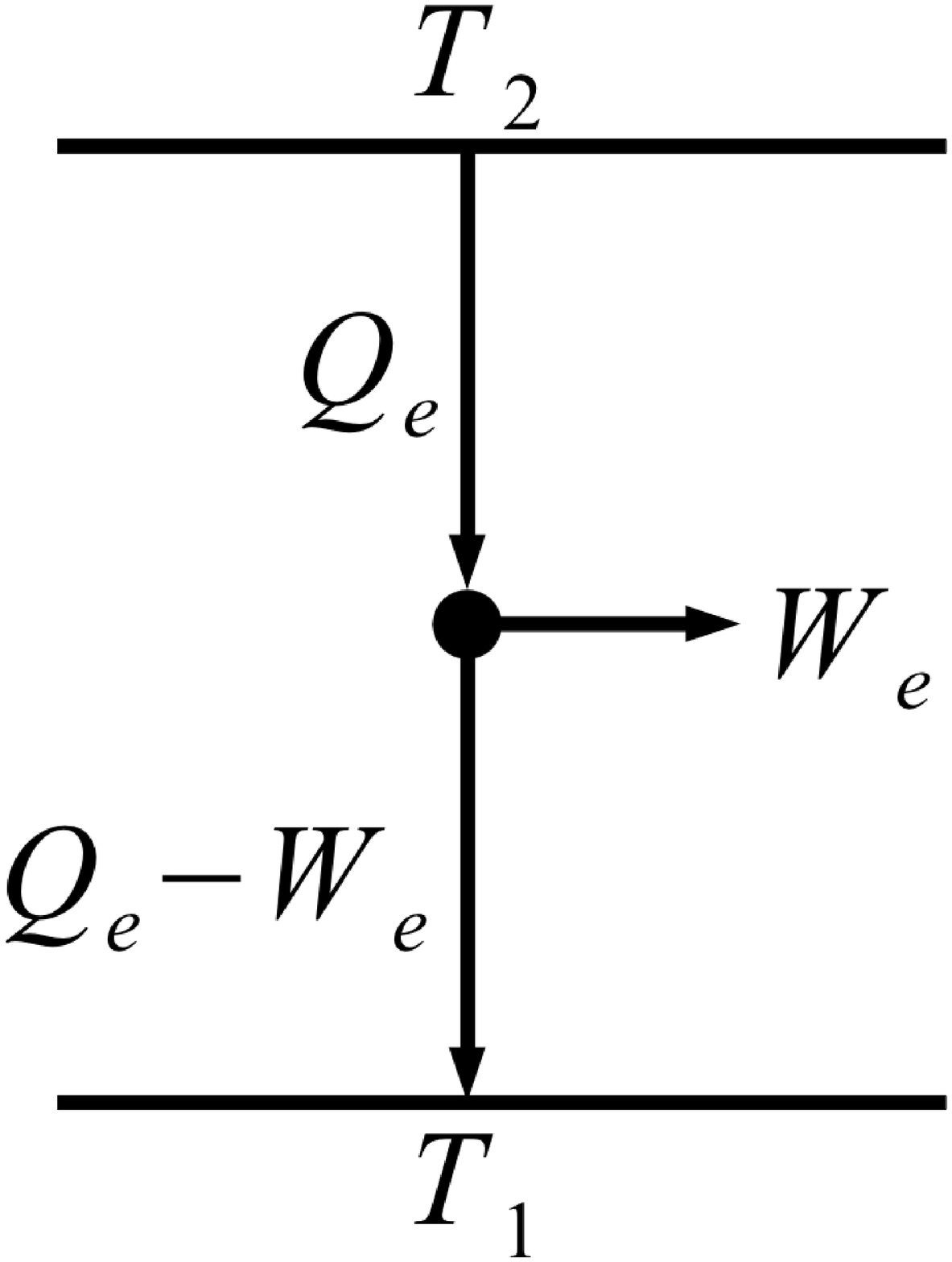}}
            \label{fg:carnotengine}
    }
    \caption{Reliable Heat Pumps and Engines\label{fg:carnotpumpengine}}
 \end{figure}
Demonstration of the logical equivalence of each pair of these statements can easily be found in a textbook such as \cite{Adk68}.  The equivalence is typically proven by the means of diagrams such as in Figure \ref{fg:carnotequiv}.  This diagram shows the combination of heat engine and heat pumps being used to attempt violations of the Kelvin and Clausius statements.    Figure \ref{fg:vpumpkelv} shows that if a heat pump can operate with efficiency $n_p={Q_c \over W_p} > n_{CP}={Q_c \over W_c}$, then in combination with a reversible heat engine operating at $n_{CE}={W_c \over Q_c}$ there is a net conversion of $W_c-W_p >0$ heat from the lower temperature heat bath into work, violating the Kelvin statement.  Similarly Figure \ref{fg:vengclaus} shows a heat engine operating with efficiency $n_e={W_c \over Q_e} > n_{CE}={W_c \over Q_c}$ can be combined with a reversible heat pump operating at $n_{CP}={Q_c \over W_c}$ could transfer heat$Q_c-Q_e > 0$ from a colder to hotter heat bath without requiring work, thus violating the Clausius statement.
It should be noted that this demonstration requires a number of usually unstated assumptions, such as the absence of negative temperatures.  In particular, the equivalence requires it to be physically possible to construct a reversible heat engine or pump.  For example, if it were not physically possible to build a heat engine whose efficiency could reach that of a theoretical reversible heat engine, then it would not necessarily follow that a real heat pump exceeding the Carnot efficiency could violate the Kelvin or Clausius versions of the second law\footnote{Suppose for all real heat engines $n_e \leq n_{max} < n_{CE}$. All that could be implied would be that the efficiency of real heat pumps were bounded by $n_p \leq {1 \over n_{max}}$ but ${1 \over n_{max}} > n_{CP}$.  Note that such a heat pump, with $n_p > n_{CP}$, would not be possible to operate reversibly as a heat engine.}.

The problem arises that fluctuation phenomena, such as Brownian motion, do, in principle, violate all these statements of the second law, when probabilistic processes are allowed.  Attempts to define a modified second law will typically restrict the domain of validity.  It may be suggested that the second law only applies to the thermodynamic limit of an infinite number of atoms, where fluctuations become negligible, or it may be suggested that the second law only applies to continuous or reliable processes:
\begin{itemize}
\item No \textit{reliable} process is possible whose sole result is the extraction of heat from a heat bath and its conversion to work.
\item No process is possible \textit{with probability one}, whose sole result is the extraction of heat from a heat bath and its conversion to work.
\item No \textit{continuously operating} process is possible whose sole result is the extraction of heat from a heat bath and its conversion to work.
\item No process is possible whose sole result is, \textit{on average}, the extraction of heat from a heat bath and its conversion to work.
\end{itemize}

Restricting the domain of validity in this way, however, proves unable to provide answers to many interesting questions about the thermodynamic consequences of fluctuation phenomena.  Can systems with a finite number of atoms be used to continuously, reliably convert heat to work?  If a process can succeed with probability less than one, how much work can be extracted?  If a process only operates for a finite amount of time how much work can be extracted? Can it be arbitrarily large?  Can a process exist which can extract an arbitrarily large quantity of work with probability arbitrarily close to one, while still failing on average due to catastrophic failure when it does fail?

This can be illustrated by considering a hypothetical family of processes, parameterised by $N>1$. Process $N$ will, with probability $1-{1 \over N}$, generates $N$ units of work from heat, but with probability $1 \over N$ it requires $N^2$ units of work to be dissipated.  The mean work produced is $-1$, regardless of the value of $N$, but as $N \rightarrow \infty$ arbitrarily large amounts of work are produced with probability arbitrarily close to one.  Even more extreme examples can easily be constructed.  Such a family of processes satisfies several of the restricted laws above, but does not accord with our experience of fluctuation phenomena.

\section{Fluctuations and the Second Law}\label{s:fluctuation}
In this Section the main argument of the paper will be explored.  Rather than follow the path of the modifications in Section \ref{s:phenom}, restricting the domain of validity of the second law so as to exclude fluctuation phenomena, it will instead be expanded to include fluctuation phenomena.  Fluctuations will be treated as being probabilistic processes, occurring with probability less than one.  The modified law should set a constraint upon the size of fluctuations that can occur, and should reduce to the fluctuation-free second law when only deterministic processes occur.

The proposed modification to the phenomenological second law is based upon nothing more than the observation that the greater the size of the fluctuation, the less probable its occurrence.  From this it is proposed that, for a given size of fluctuation, there is a maximum possible likelihood of it occurring:
\begin{quote}
There is no cyclic process\footnote{When discussing probabilistic cycles, a cyclic process will mean a process which returns to its original state with probability $p$, but with probability $1-p$ may end up in a different state to its starting point.}, whose sole result is the extraction of a quantity of heat, $Q$, from a heat bath at temperature $T$, and its conversion to work, which can occur with probability $p$, unless:
\begin{equation}
p \leq f(Q,T)
\end{equation}
\end{quote}
where $f$ is a function whose properties will be deduced from internal consistency.  The definition is such that it is assumed for any given $Q$ and $T$ there exists an actual physical processes which can get arbitrarily close to occurring with probability $f(Q,T)$.  If not, then there must exist a lower value of $f$ that should have been used instead.

It is possible to immediately note some properties of $f$: as the function bounds a probability, it cannot become negative; it is always possible to dissipate work as heat; if there is a process that extracts $Q^\prime >Q$ with probability $p$, then by also dissipating work $W=Q^\prime-Q$, there is a process that extracts $Q$ with probability $p$.  These immediately constrain the function:
\begin{eqnarray}
f(Q,T) & \geq &0 \\ f(Q,T) &=&1 \;\; \forall Q \leq 0 \\  f(Q,T) &\geq & f(Q^\prime,T) \;\; \forall Q^\prime > Q
\end{eqnarray}
The last condition implies that if $f$ is also a differentiable function of $Q$, then
\begin{equation}
 \partxy{f}{Q}  \leq 0  \;\; \forall Q
\end{equation}
One trivial solution would be: $f(Q,T)=0, \;\; \forall Q>0$.  This would correspond to all fluctuations being forbidden.  At the other extreme, $f(Q,T)=1, \;\; \forall Q$ would imply one could get arbitrarily close to any size of fluctuation, at any probability.

This is a more restrictive condition than the mean conversion of heat to work over cycle being negative, although it does imply it. The proof of this is straightforward. If there exists a process which can produce a positive expectation value for production of work over a single cycle, then repeating that cycle a large number of times produces an expectation value as large as one likes, with a gaussian spread around that mean.  The probability that any given quantity of work can be exceeded becomes close to one.  Hence any process which can produce a positive expectation value for work will, on repeated application, exceed any function $f<1$.

This kind of fluctuation - extracting work from a single heat bath - will be called a Kelvin fluctuation, and be represented as in Figure \ref{fg:kelvin}, showing $W$ work being extracted from a heat bath at temperature $T_1$.
\doublepict{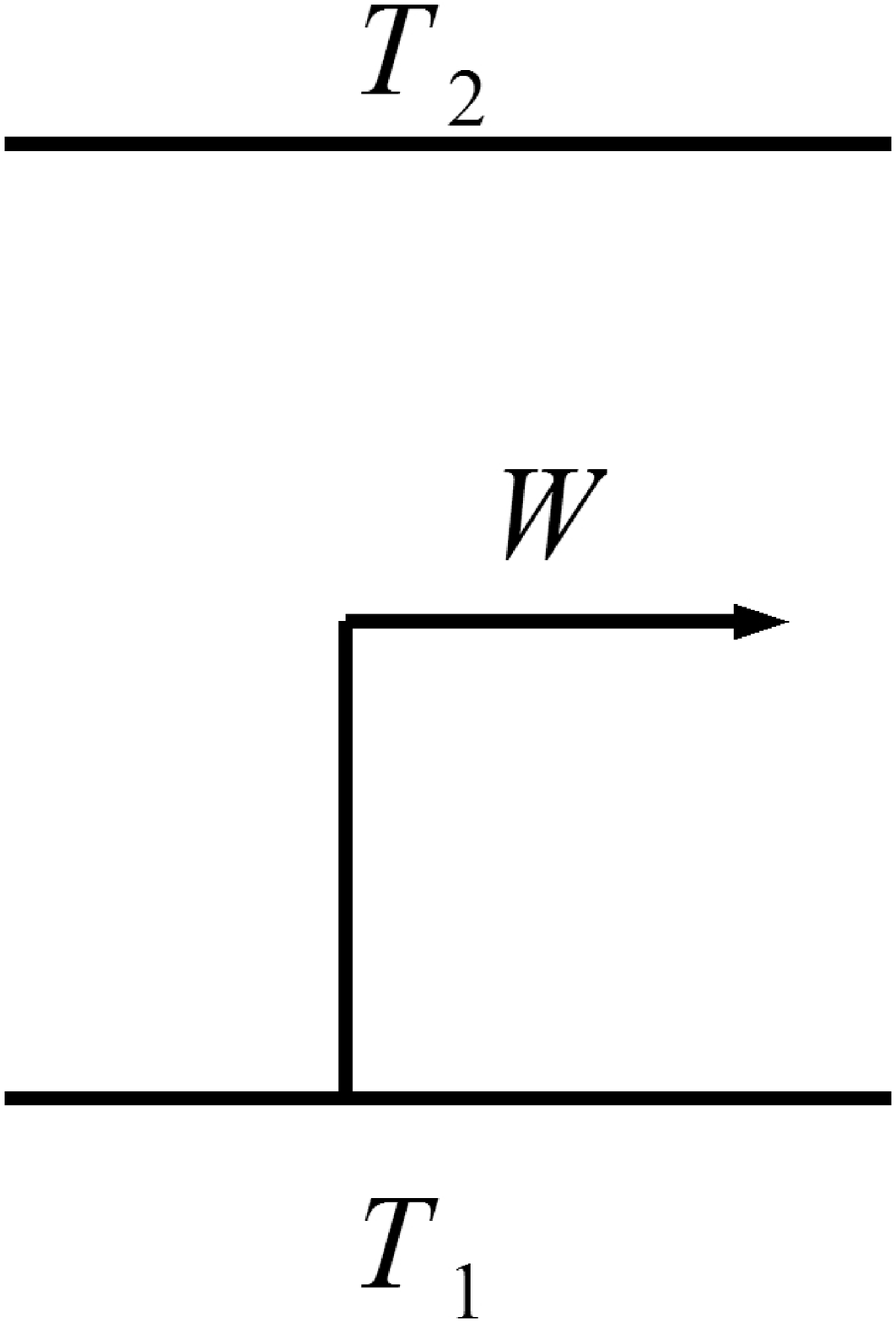}{kelvin}{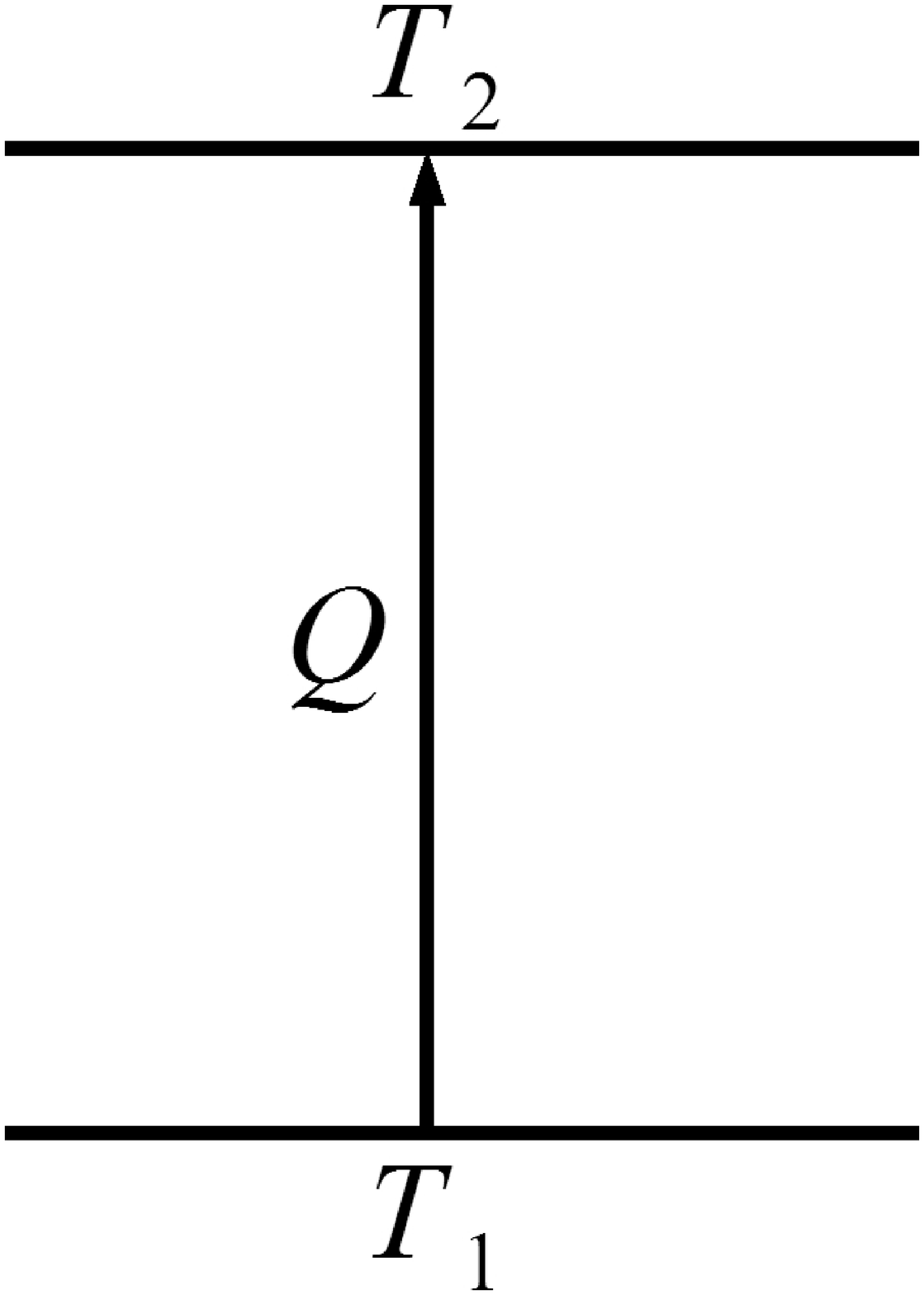}{clausius}{Kelvin and Clausius Fluctuations}{kelvinclausius}{0.2}

The equivalence of Kelvin fluctuations to other kinds of fluctuations will now be demonstrated.
\subsection{Kelvin and Clausius Fluctuations}\label{ss:kelvtoclaus}
A Clausius fluctuation, as in Figure \ref{fg:clausius}, will denote the spontaneous transfer of $Q$ work from a heat bath at $T_1$ to a heat bath at $T_2 > T_1$ occurring with a maximum probability $f_C(Q,T_1,T_2)$.  One way to achieve a Clausius fluctuation is given in Figure \ref{fg:kelvtoclaus}, combining a Kelvin fluctuation with a reliable Carnot pump operating at efficiency $n_{CP}={Q \over W}={T_2 \over {T_1- T_2}}$.  This can occur with probability $f(W,T_1)$, so  $f_C(Q,T_1,T_2)$ cannot be less than this: $f_C(Q,T_1,T_2)\geq f(W,T_1)=f({Q \over n_{CP}},T_1)$.
\doublepict{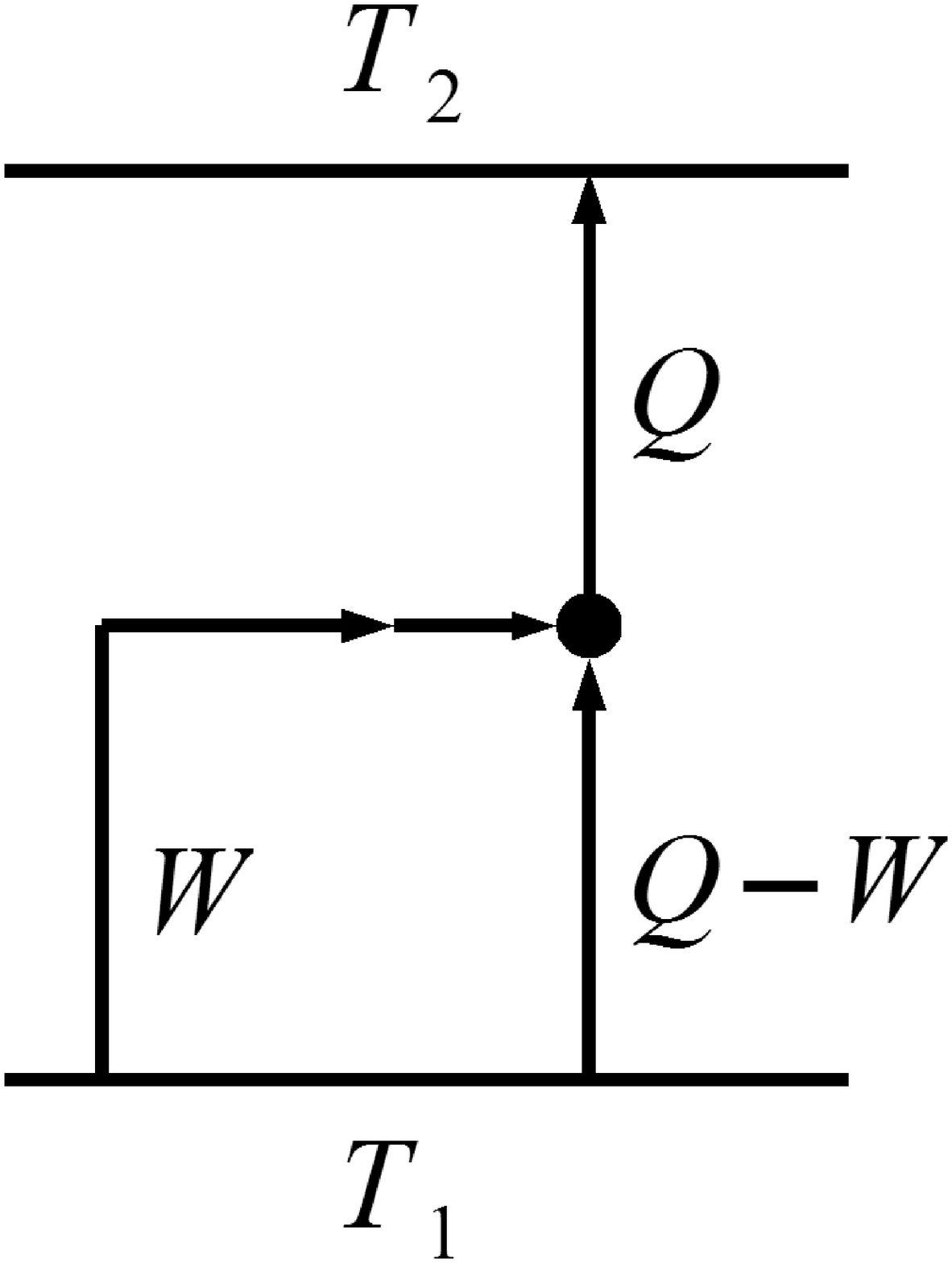}{kelvtoclaus}{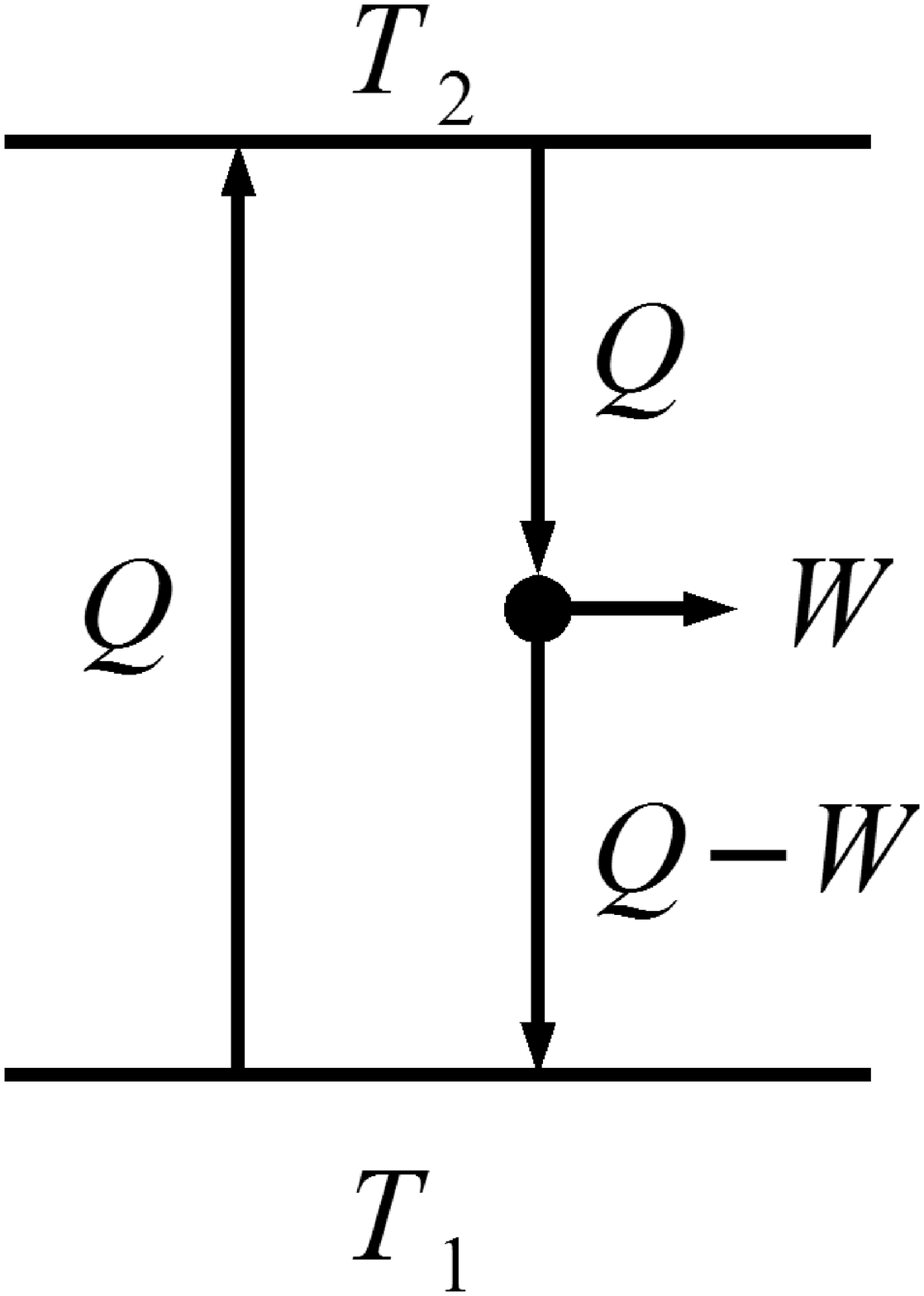}{claustokelv}{Converting Kelvin and Clausius Fluctuations}{kelvintoclausius}{0.2}
A Kelvin fluctuation can similarly (Figure \ref{fg:claustokelv}) be created from a Clausius fluctuation, by allowing the heat $Q$ from the Clausius fluctuation to drive a reliable Carnot engine at efficiency $n_{CE}={W \over Q}=1- {T_1 \over T_2}$.  This implies  $f_C(Q,T_1,T_2)\leq f(W,T_1)=f(Q n_{CE},T_1)$ and $n_{CE}= {1 \over n_{CP}}$ establishes
\begin{eqnarray}
f_C(Q,T_1,T_2)&=&f\left({Q \over n_{CP}},T_1\right) =f\left({Q n_{CE}},T_1\right) \nonumber \\
 &=&f\left(Q\left(1-{T_1 \over T_2} \right),T_1\right)
\end{eqnarray}

 \begin{figure}[htb]\centering
    \subfigure[$n_{P}={Q \over W} > n_{CP}$ ] {
       \resizebox{0.2\textwidth}{!}{\includegraphics{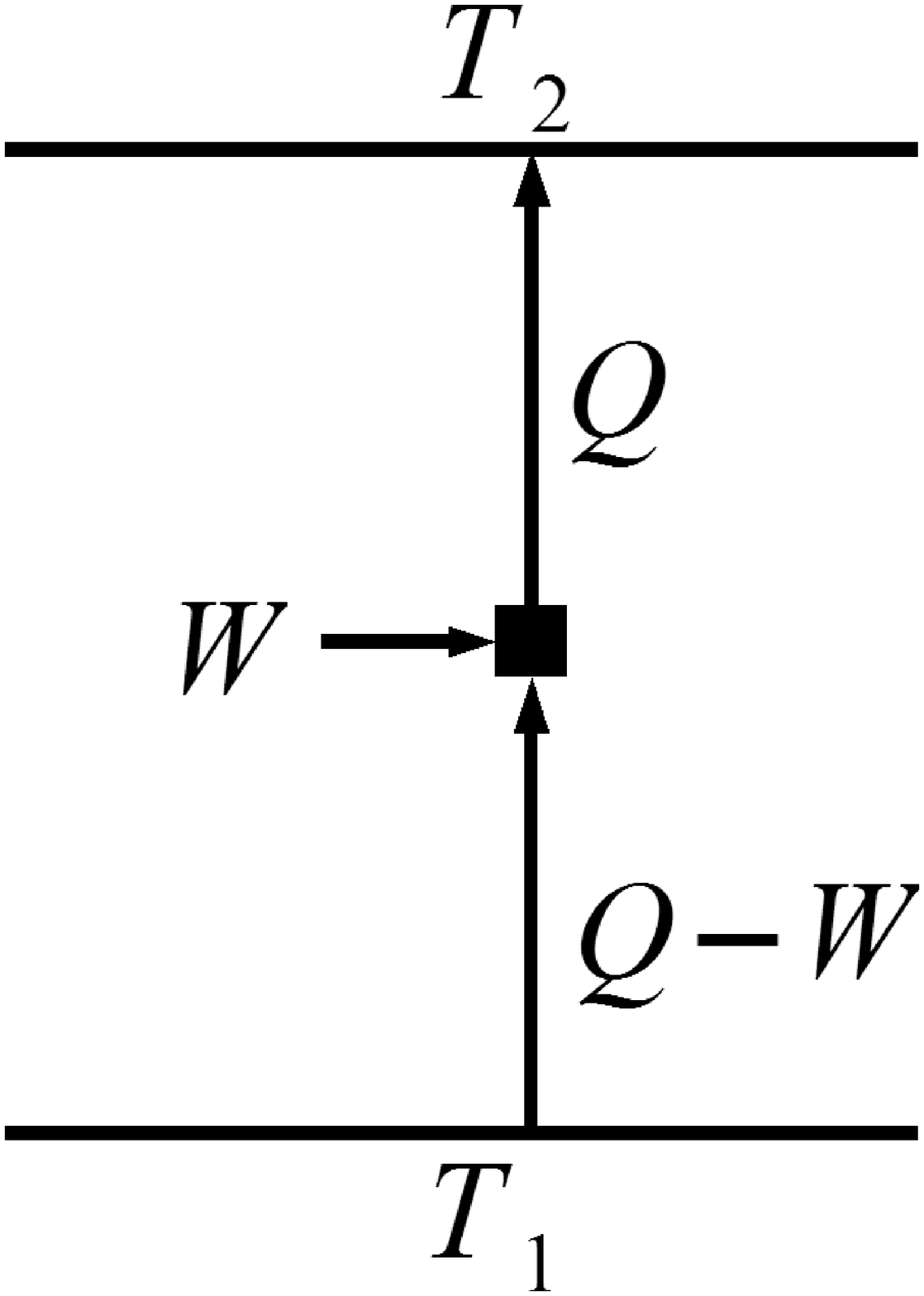}}
            \label{fg:pump}
    }
    \subfigure[ $n_{E}={W \over Q} > n_{CE}$ ]{
           \resizebox{0.2\textwidth}{!}{\includegraphics{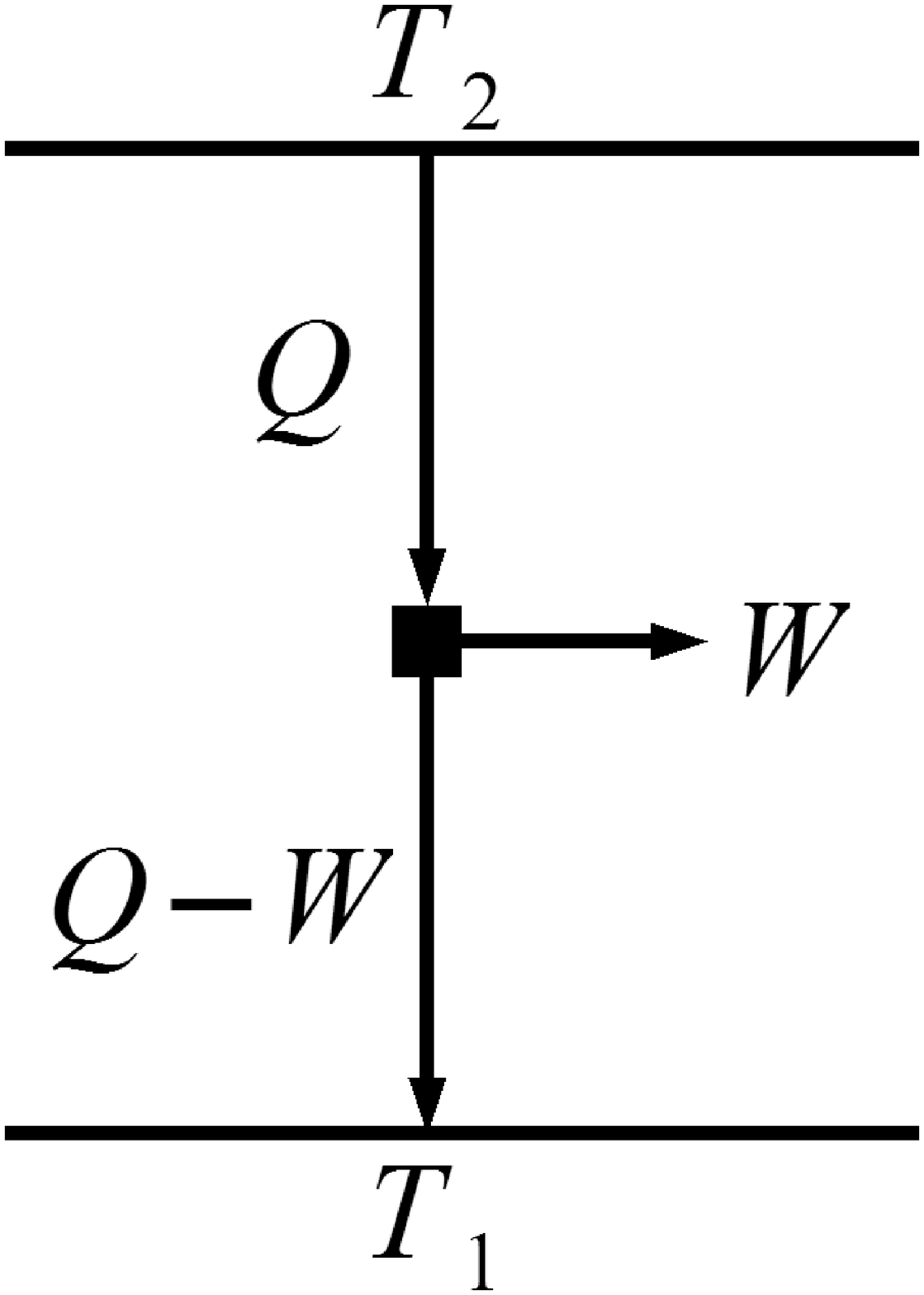}}
        \label{fg:engine}
    }
    \caption{Fluctuation Heat Pumps and Engines \label{fg:flucpumpengine}}
 \end{figure}

\subsection{Kelvin, Clausius and Heat Pump  Fluctuations}\label{ss:kelvclaustopump}
\begin{figure}[htb]\centering
    \subfigure[ ] {
       \resizebox{0.2\textwidth}{!}{\includegraphics{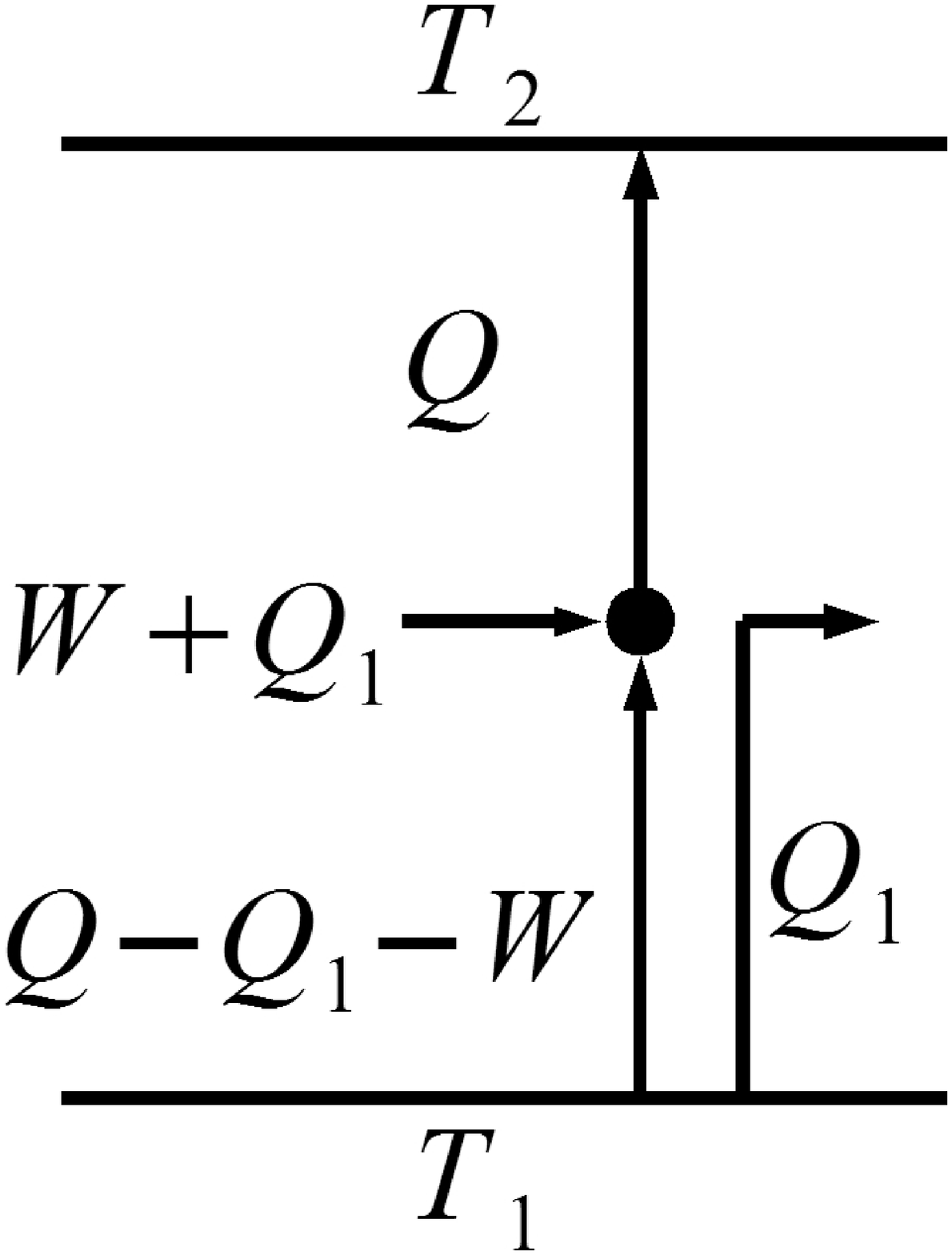}}
            \label{fg:kelvtopump}
    }
    \subfigure[  ]{
           \resizebox{0.2\textwidth}{!}{\includegraphics{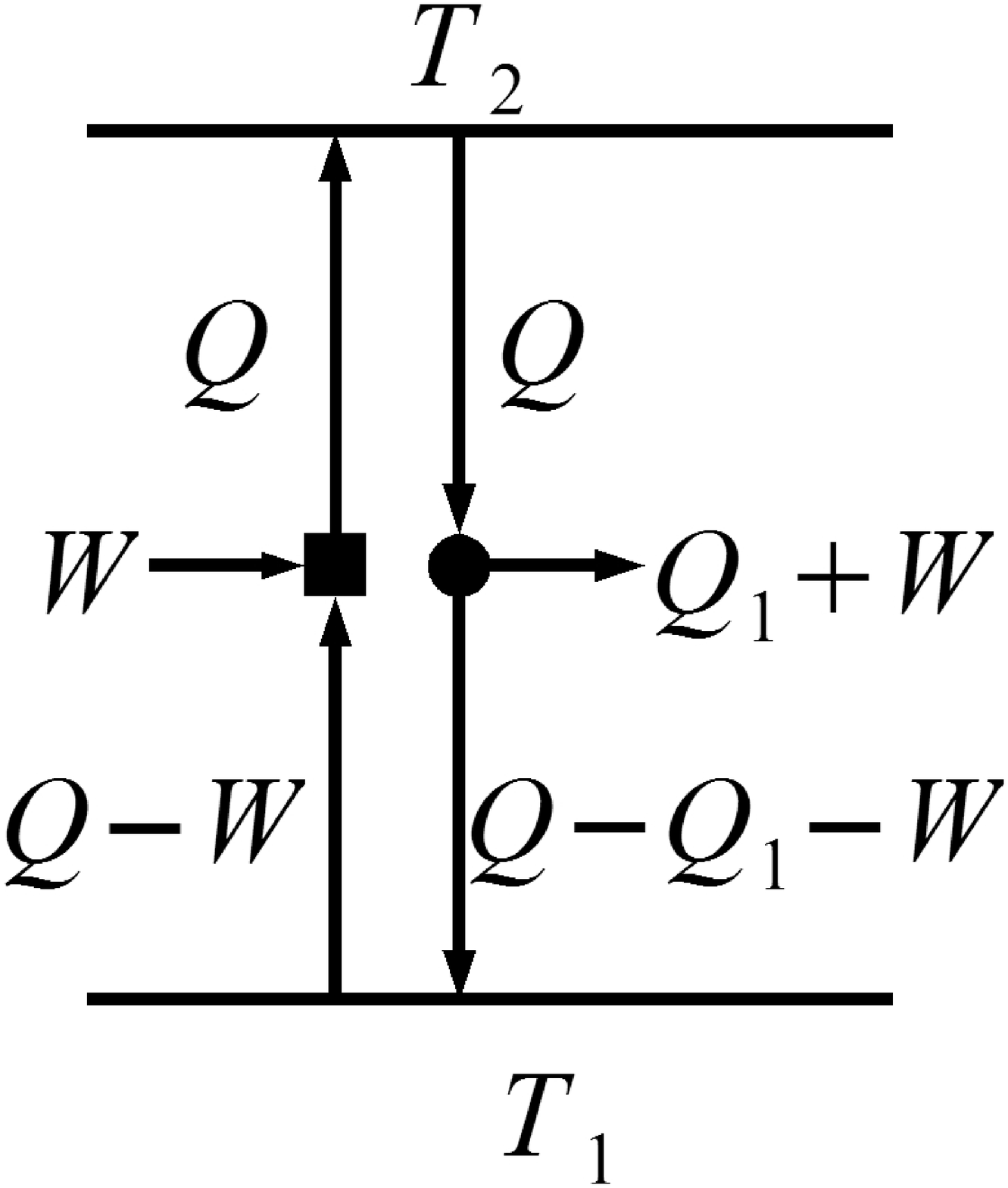}}
        \label{fg:pumptokelv}
    }
    \caption{Kelvin Fluctuations and Fluctuation Heat Pumps \label{fg:kelvintopump}}
 \end{figure}

A fluctuation heat pump (Figure \ref{fg:pump}) is a heat pump that is able to operate with a higher efficiency than a reversible Carnot heat pump, but only with a probability less than one of success.
The maximum probability of success, $f_P(W,n_P,T_1,T_2)$ of achieving efficiency $n_P={Q \over W} > n_{CP}$ can be deduced either from the Kelvin fluctuation law (Figure \ref{fg:kelvintopump}) or the Clausius fluctuation law (Figure \ref{fg:clausiustopump}).
In the Figure \ref{fg:kelvtopump}, creating a fluctuation pump with efficiency $n_{P}={Q \over W} > n_{CP}$, by augmenting the behaviour of a regular Carnot pump with a Kelvin fluctuation shows $f_P(W,n_P,T_1,T_2)\geq f(Q_1,T_1)$ .  In Figure \ref{fg:pumptokelv}, creating a Kelvin fluctuation of size $Q_1$, by extracting the heat pumped by fluctuation heat pump at efficiency $n_{P}={Q \over W} > n_{CP}$, and using it to drive a Carnot heat engine gives $f(Q_1,T_1) \geq f_P(W,n_P,T_1,T_2)$.  Substituting $Q_1 n_{CP}=W(n_P-n_{CP})$ gives
\begin{equation}
f_P(W,n_P,T_1,T_2)=f\left(W\left({n_P \over n_{CP} } -1 \right),T_1\right)
\end{equation}
\doublepict{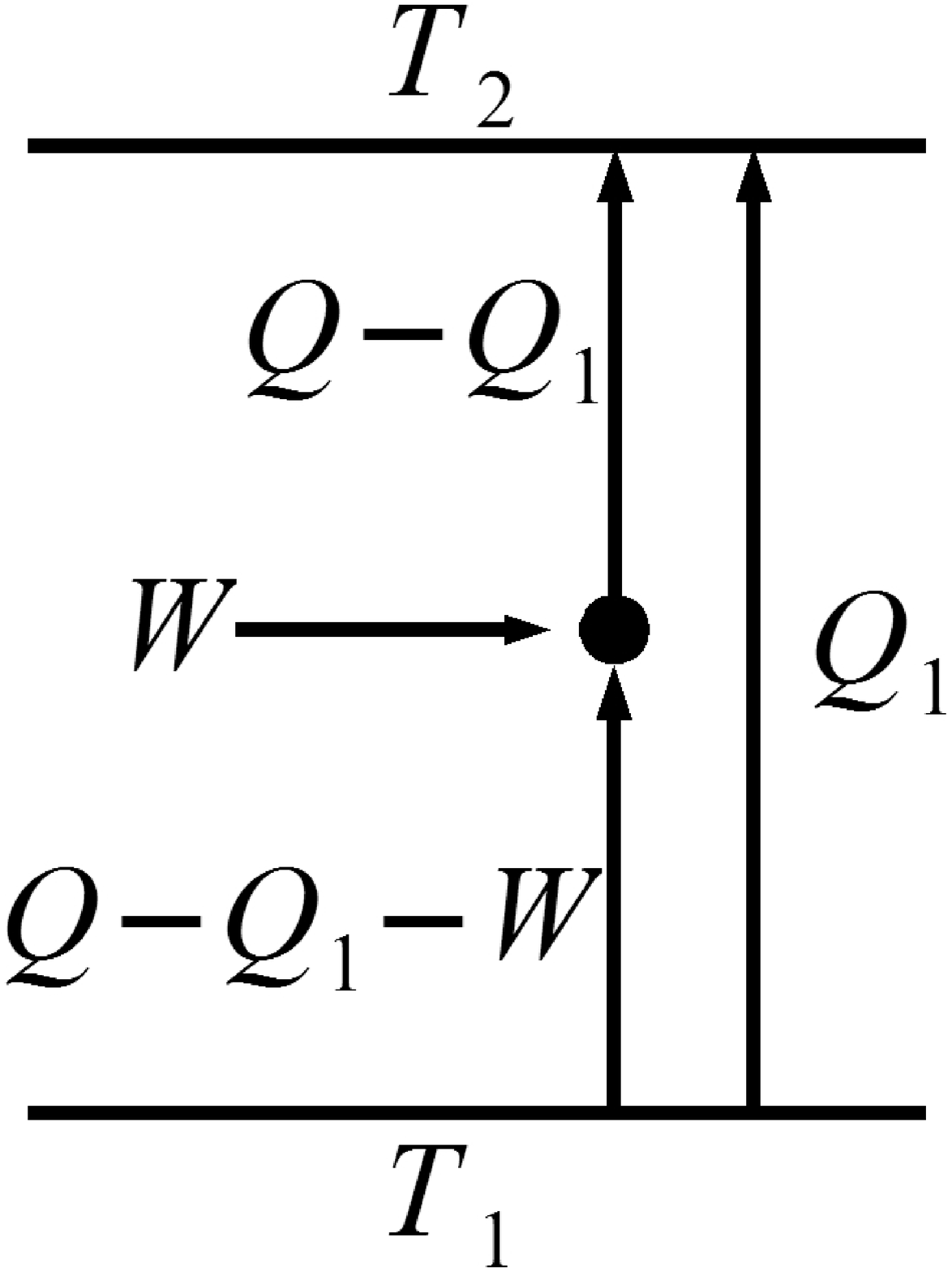}{claustopump}{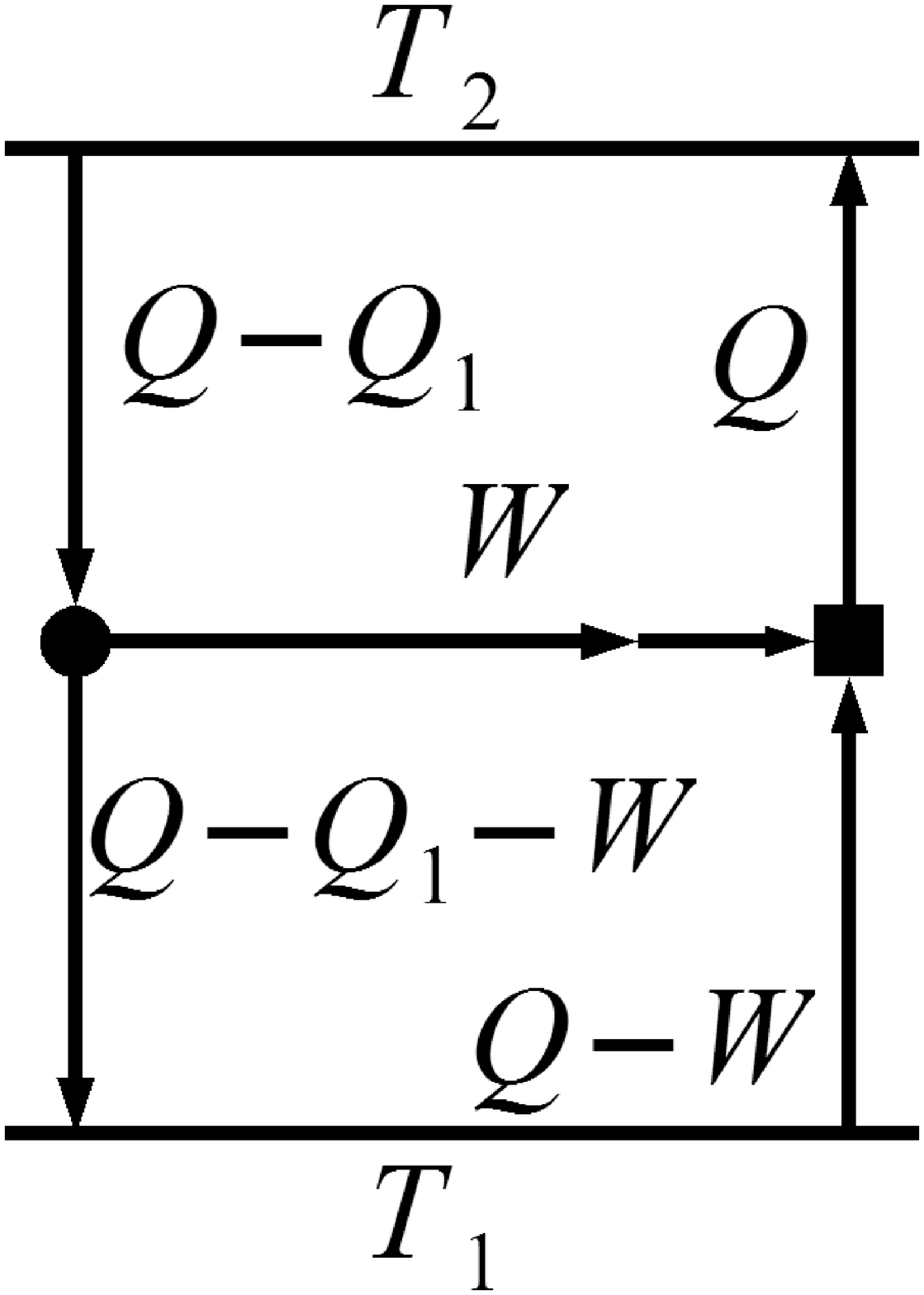}{pumptoclaus}{Clausius Fluctuations and Fluctuation Heat Pumps}{clausiustopump}{0.2}
Figure \ref{fg:claustopump} augments the Carnot heat pump with a Clausius fluctuation of size $Q_1$ to create a fluctuation pump of efficiency $n_P={Q \over W} > n_{CP}$.  Now using the work extracted from a Carnot engine to drive a fluctuation heat pump, gives a Clausius fluctuation in Figure \ref{fg:pumptoclaus}.  Combined $f_P(W,n_P,T_1,T_2)=f_C(Q_1,T_1,T_2)$ with $Q_1=W(n_P-n_{CP})$, so
\begin{equation}
f_P(W,n_P,T_1,T_2)=f_C\left(W(n_P - n_{CP}),T_1,T_2\right)
\end{equation}
It can be easily confirmed that this is consistent with the relationship $f_C(Q,T_1,T_2)=f({Q \over n_{CP}},T_1)$.
\subsection{Kelvin, Clausius and Heat Engine Fluctuations}\label{ss:kelvclaustoeng}
Similarly, a fluctuation heat engine (Figure \ref{fg:engine}) is a heat engine that can operate with a higher efficiency than a reversible Carnot heat engine, but only with a probability less than one of success.

Augmenting a Carnot heat engine with a Kelvin fluctuation of size $Q_1$, Figure \ref{fg:kelvtoeng}, creates a fluctuation heat engine, while using the heat pumped by a regular Carnot pump to drive a fluctuation heat engine, Figure \ref{fg:engtokelv}, creates an equivalent Kelvin fluctuation.
 \doublepict{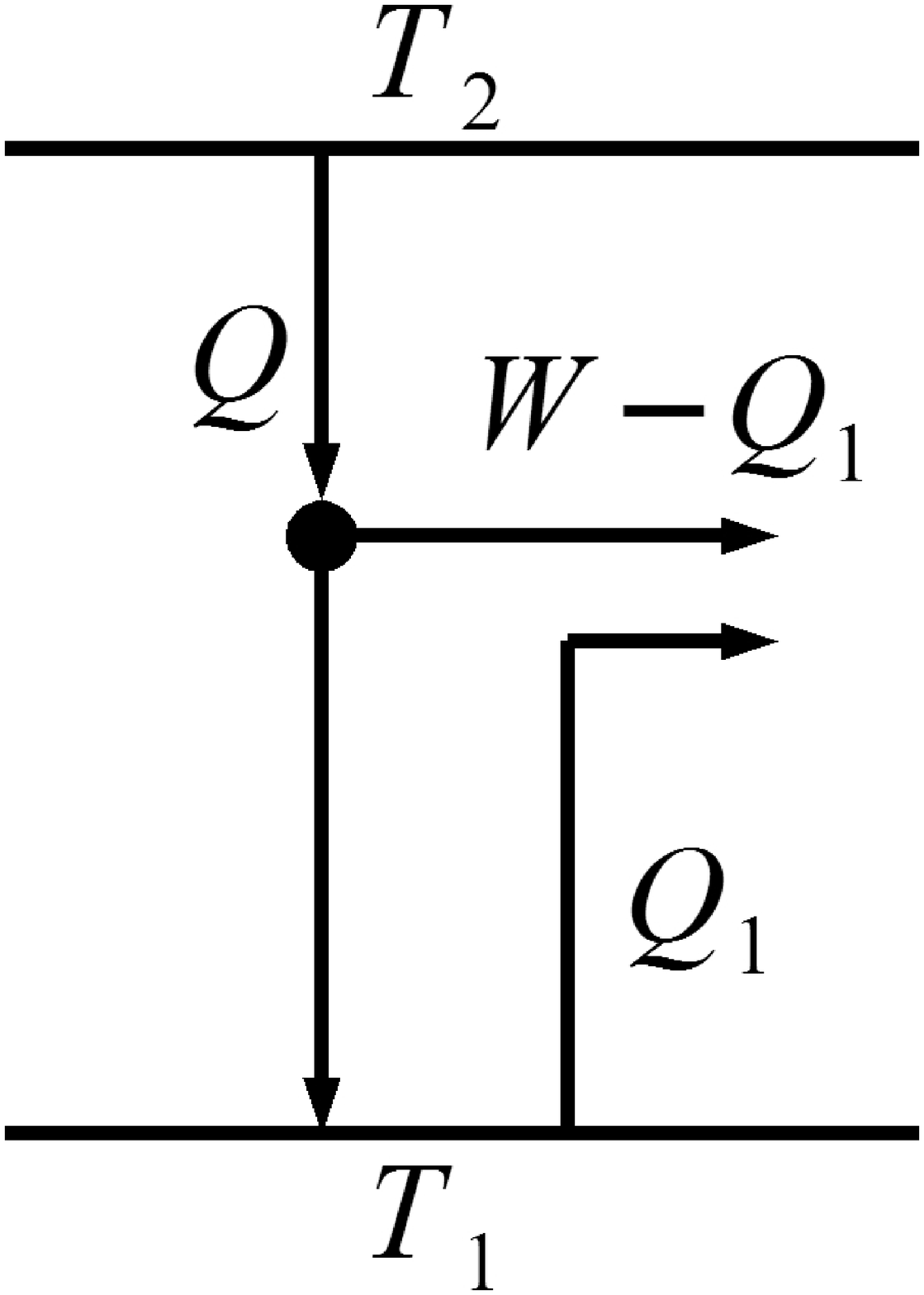}{kelvtoeng}{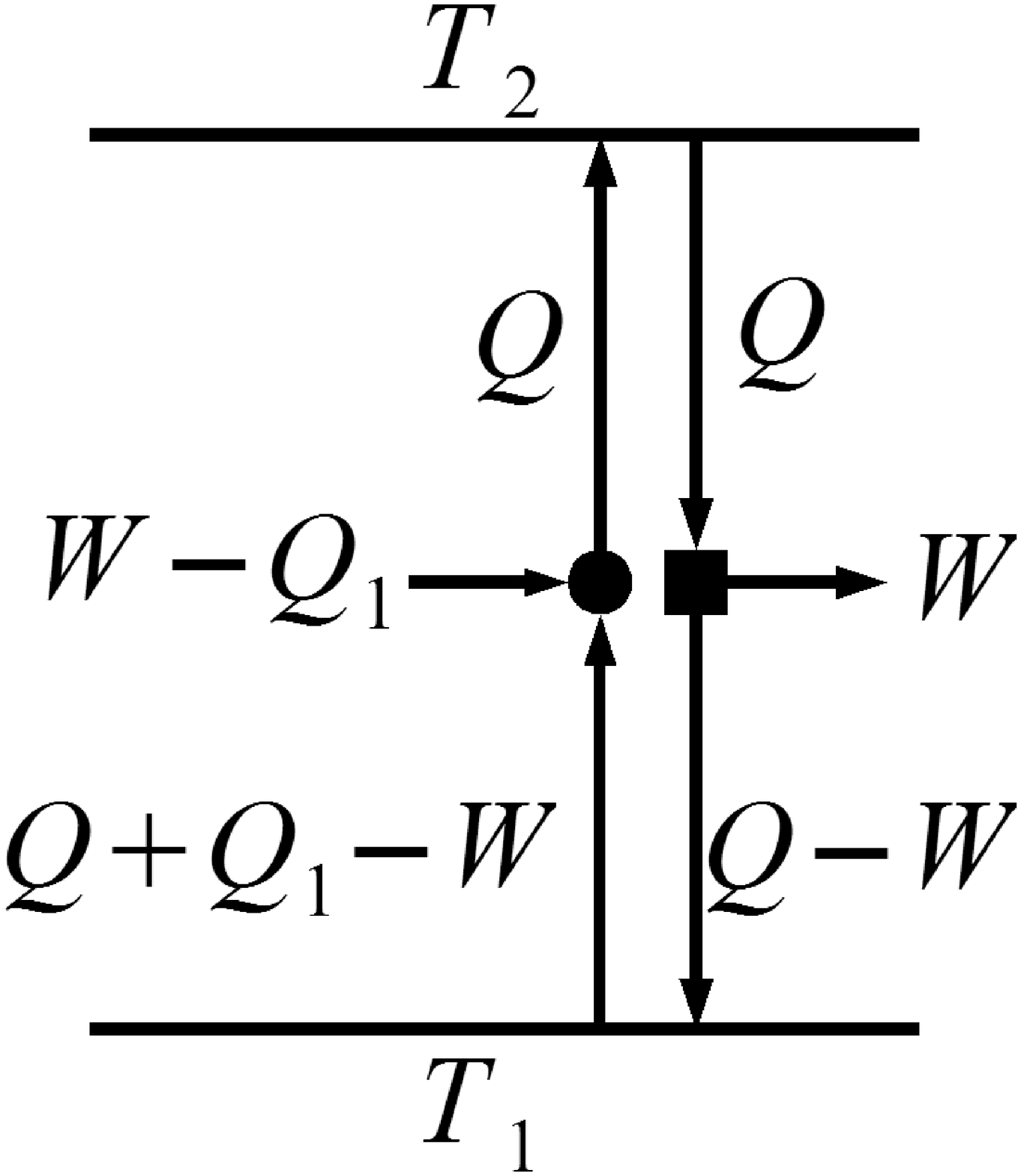}{engtokelv}{Kelvin Fluctuations and Fluctuation Heat Engines}{kelvintoengine}{0.2}
Giving the maximum probability achievable for a fluctuating heat engine to extract heat $Q$ from a heat bath at temperature $T_2$, with efficiency $n_E={W \over Q} > n_{CE}$, depositing the remainder in a heat bath at temperature $T_1 < T_2$ as $f_E(Q,n_E,T_1,T_2)$, the diagrams quickly yield $Q_1=Q(n_E-n_{CE})$ and the relationship
\begin{equation}
f_E(Q,n_E,T_1,T_2)=f(Q(n_E-n_{CE}),T_1)
\end{equation}
Figure \ref{fg:clausiustoengine} provides the equivalent analysis for Clausius fluctuations, now creating a Clausius fluctuation by driving a regular Carnot pump with the work extracted by a fluctuation heat engine.
\doublepict{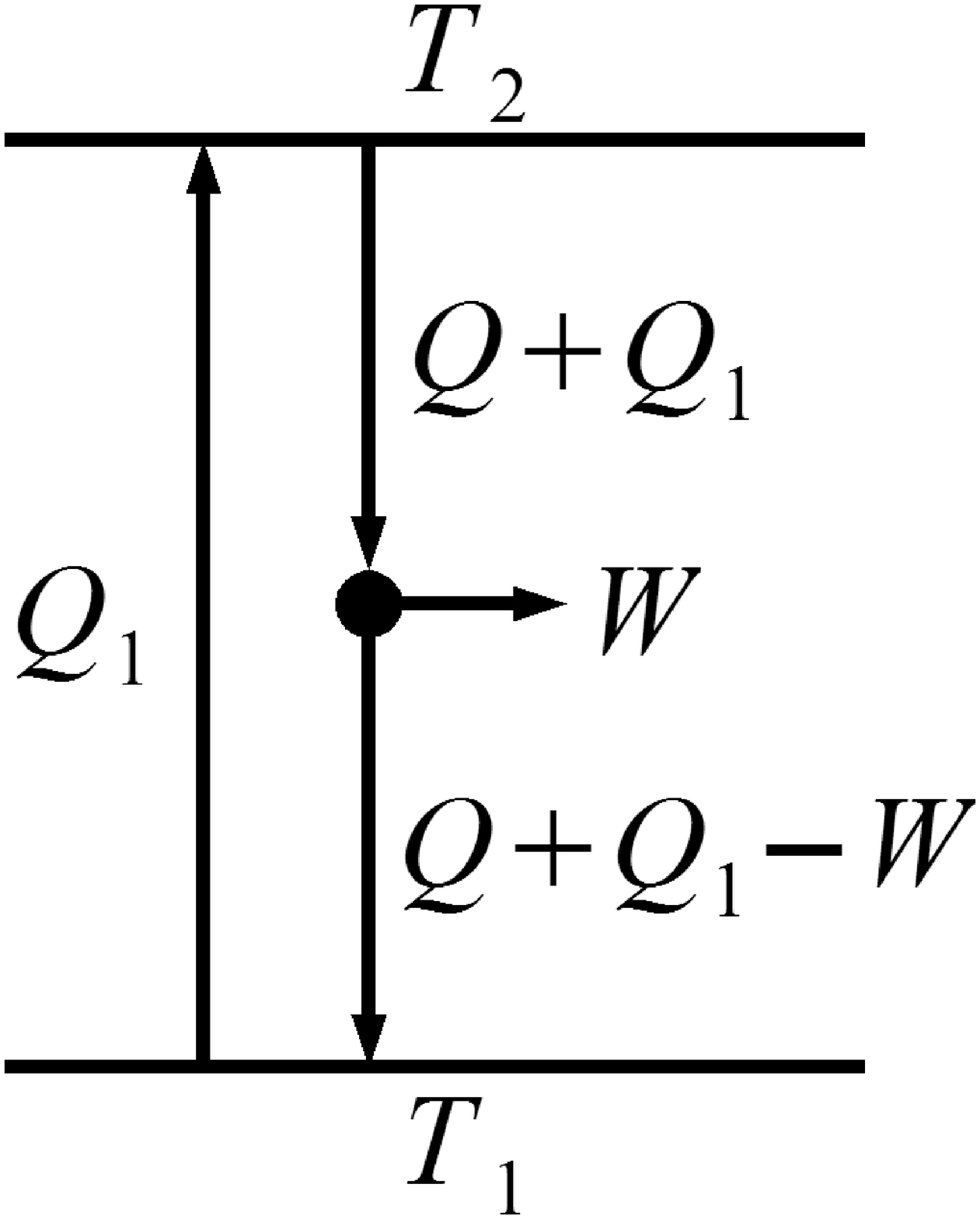}{claustoeng}{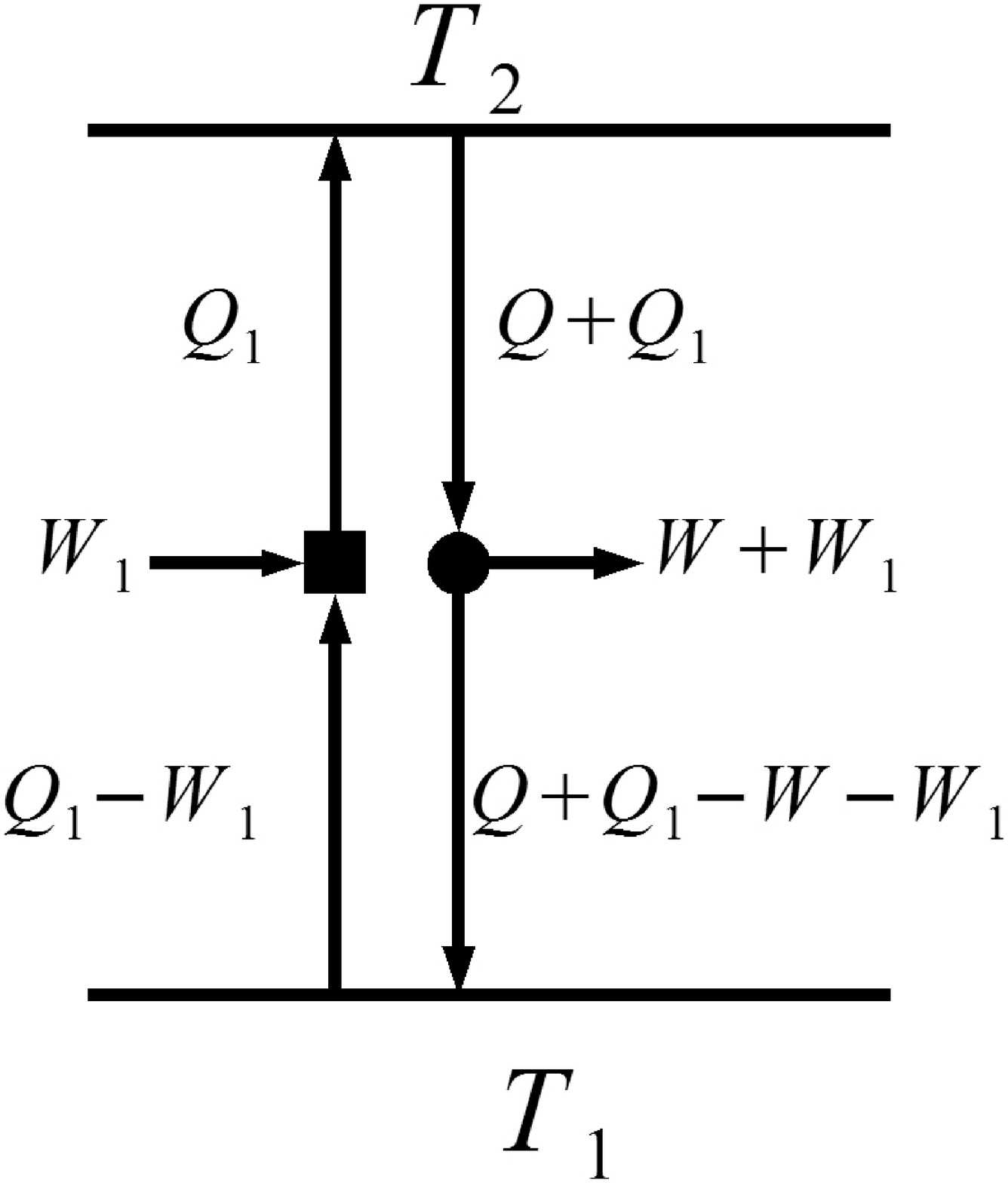}{engtoclaus}{Clausius Fluctuations and Fluctuation Heat Engines}{clausiustoengine}{0.2}
As $Q_1 n_{CE}=Q(n_E-n_{CE})$
\begin{equation}
f_E(Q,n_E,T_1,T_2)=f_C\left(Q\left({n_E \over n_{CE}}-1\right),T_1,T_2\right)
\end{equation}
Again, this is consistent with the relationship between $f_C$ and $f$.
\subsection{Heat Pumps and Engines}\label{ss:pumpeng}  It is now possible to compare the expressions for $f_E(Q,n_E,T_1,T_2)$ and $f_P(W,n_P,T_1,T_2)$ directly.  This gives $f_E(Q,n_E,T_1,T_2)=f_P(W,n_P,T_1,T_2)$ if $W\left({n_P \over n_{CP}}-1\right)=Q\left(n_E-n_{CE}\right)$. To confirm consistency this can also be derived from the diagrams in Figure \ref{fg:enginetopump}.
\doublepict{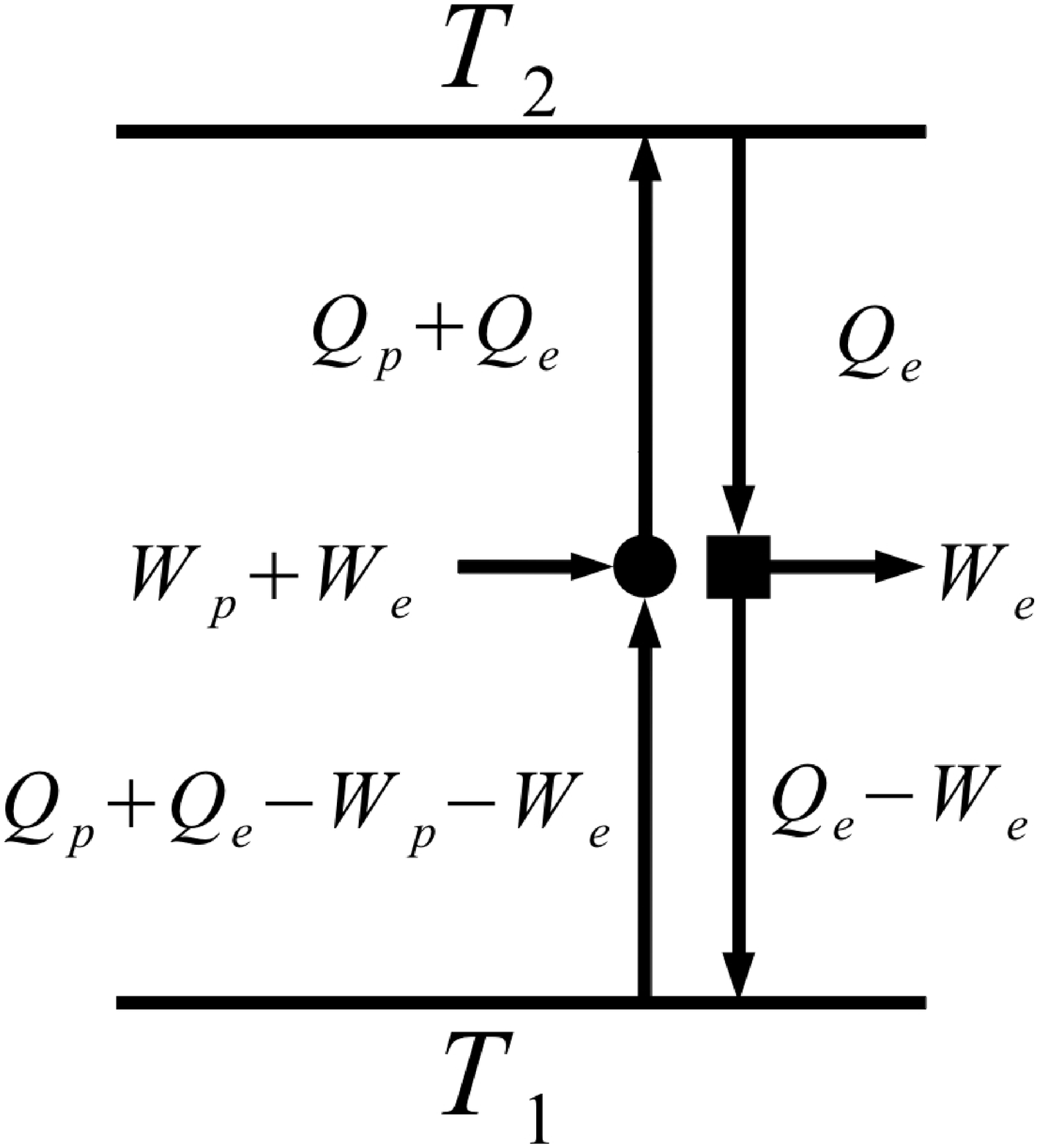}{engtopump}{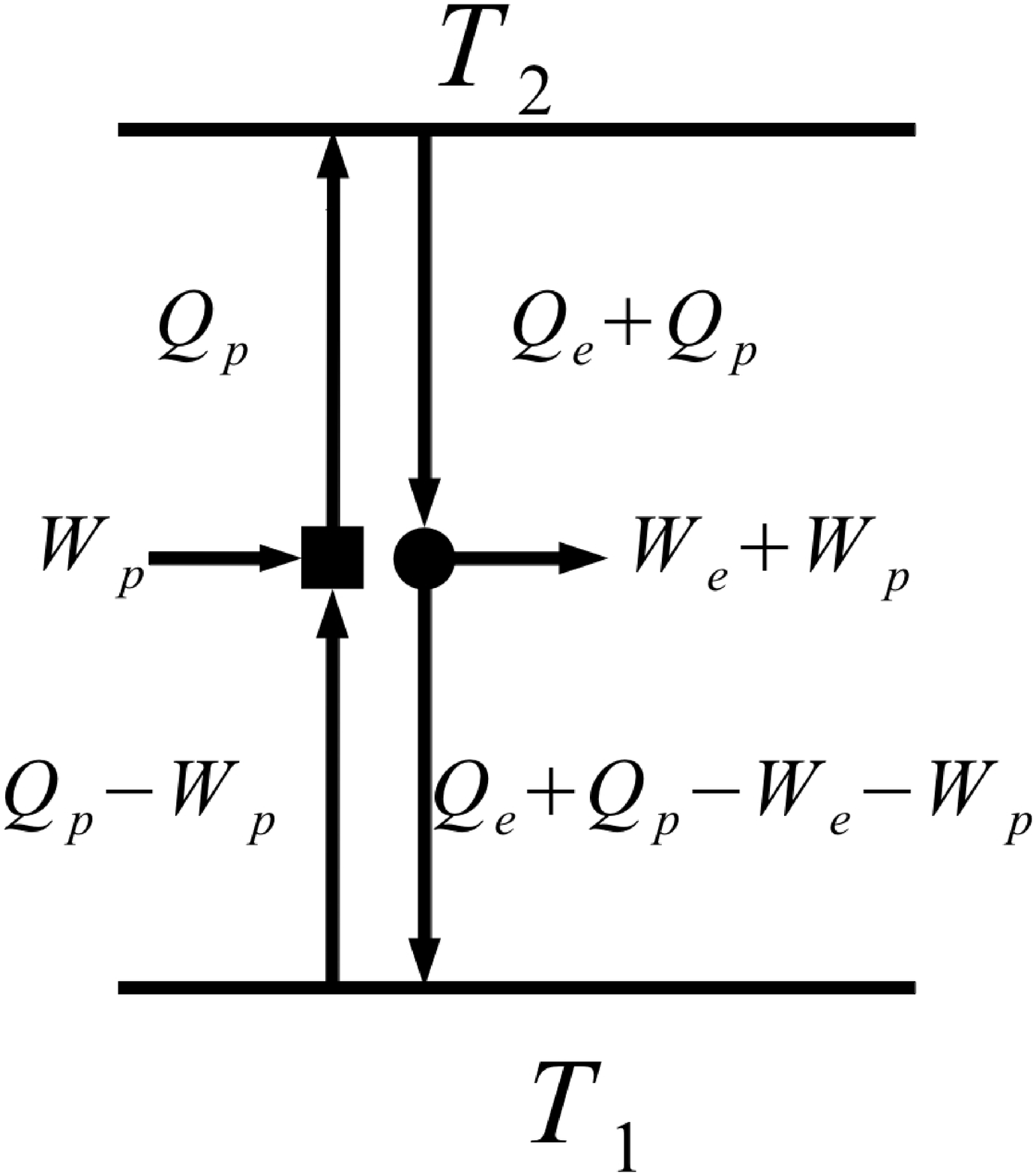}{pumptoeng}{Fluctuation Heat Pumps and Engines}{enginetopump}{0.2}
In Figure \ref{fg:engtopump}, a fluctuation heat engine, operating at $n_E={W_e \over Q_e}$ improves the efficiency of a Carnot heat pump, by using some of the pumped work to return a higher proportion of the heat into work, to create a fluctuation heat pump, with efficiency $n_P={Q_p \over W_p}$.  In Figure \ref{fg:pumptoeng}, a fluctuation heat pump, with efficiency $n_P={Q_p \over W_p}$ improves the efficiency of a Carnot heat engine to create a fluctuation heat engine with efficiency $n_E={W_e \over Q_e}$.  It can readily be confirmed that $W\left({n_P \over n_{CP}}-1\right)=Q\left(n_E-n_{CE}\right)$.
\subsection{Heat and Temperature}\label{ss:tempfluc}
There remains six diagrams for fluctuations involving two heat baths.  These diagrams determine the relationship between Kelvin fluctuations at different temperatures.  Figure \ref{fg:kelvintokelvin}
\doublepict{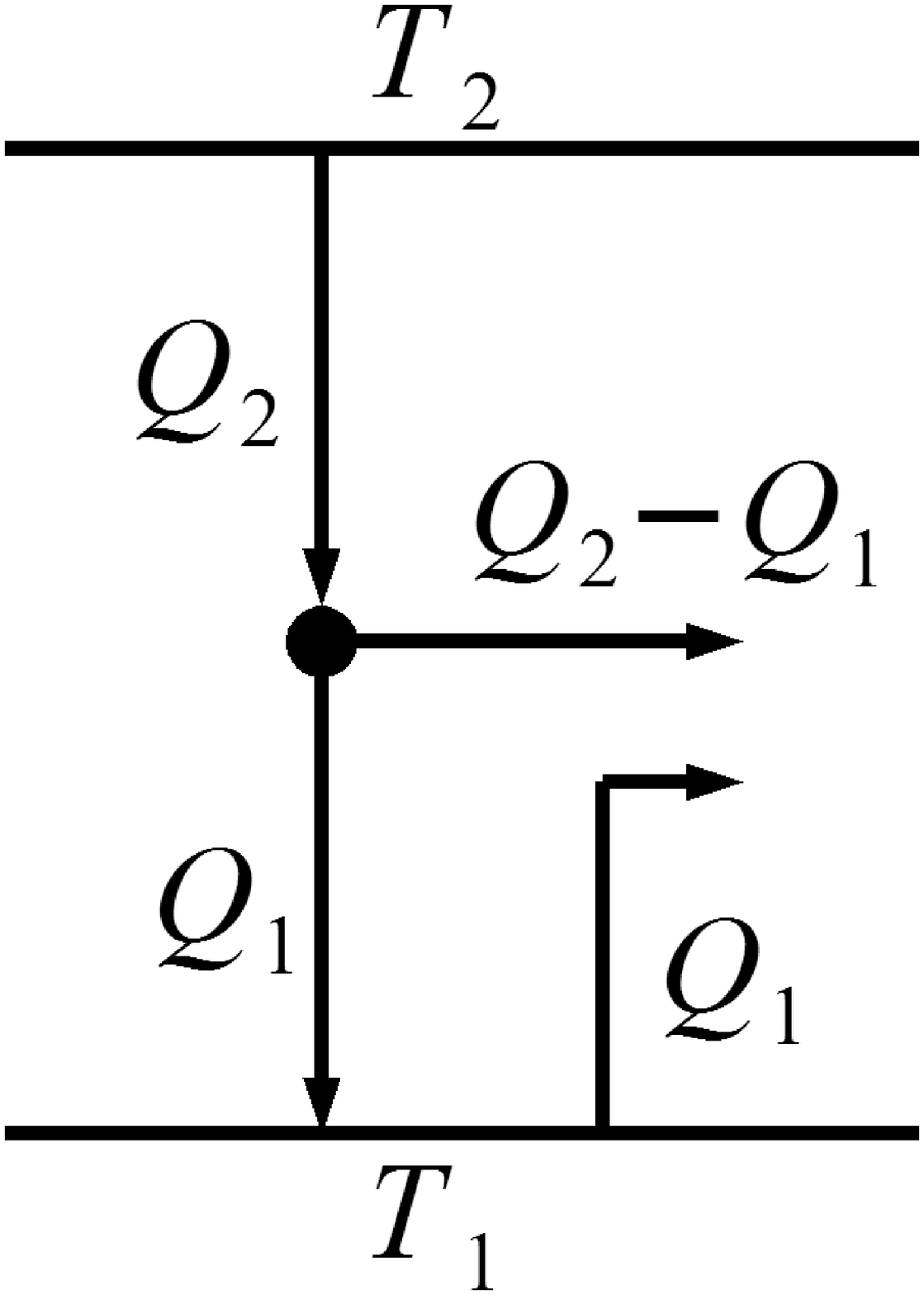}{coldtohot}{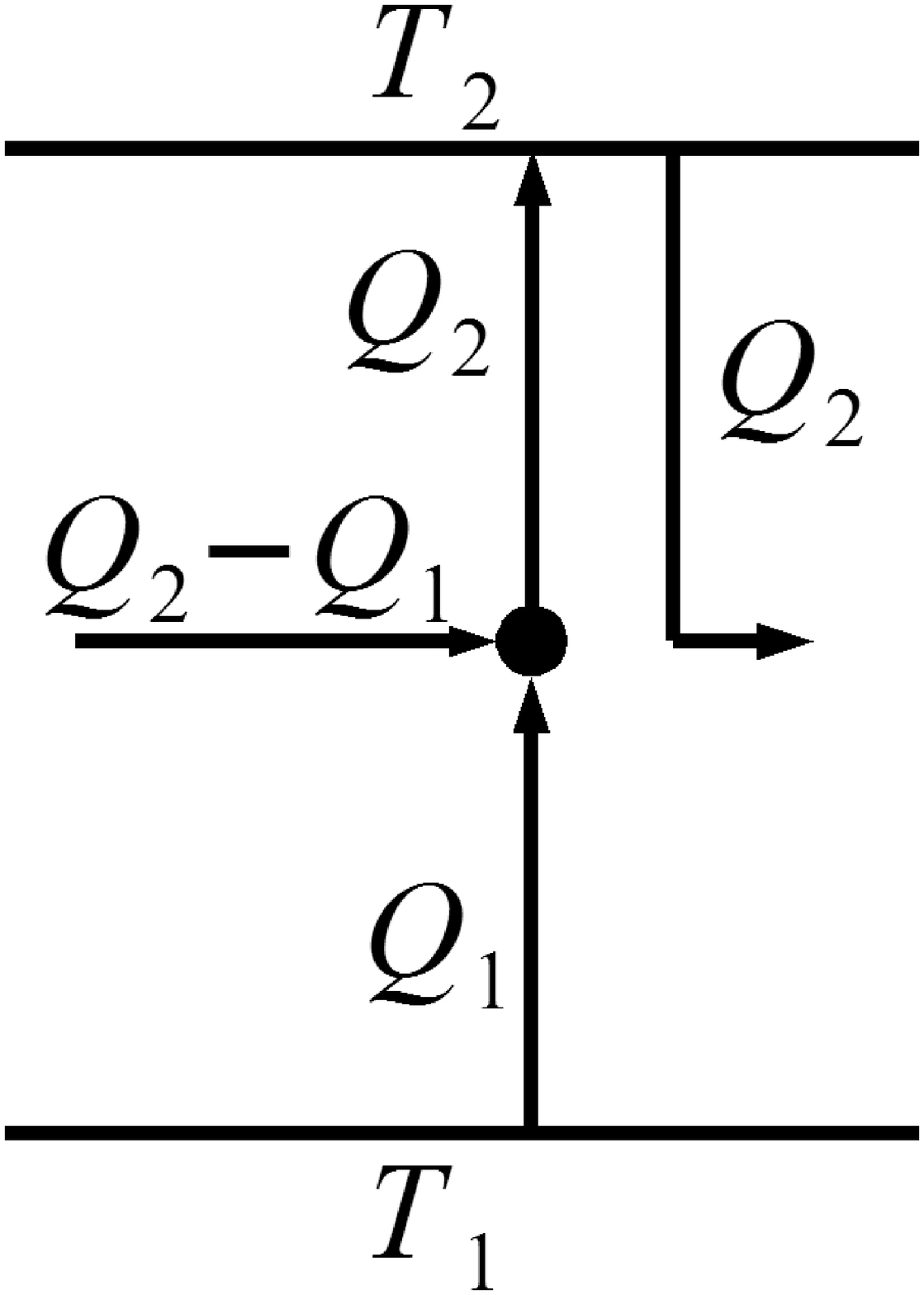}{hottocold}{Kelvin Fluctuations at Different Temperatures}{kelvintokelvin}{0.2}
shows how a Kelvin fluctuation can be converted to an equivalent Kelvin fluctuation at a higher or lower temperature, by using a Carnot pump or engine.  This supplies heat from a second bath to replace the heat obtained from the fluctuation. The overall process is then a Kelvin fluctuation from the second heat bath.

From Figure \ref{fg:coldtohot}, it can be seen that the probability of obtaining a Kelvin fluctuation of size $Q_2$ at temperature $T_2$ cannot be less that the probability of obtaining a Kelvin fluctuation of size $Q_1$ at temperature $T_1$, provided ${Q_1 \over T_1}= {Q_2 \over T_2}$.
\begin{equation}
f(Q_2,T_2) \geq f(Q_1,T_1)
\end{equation}
Figure \ref{fg:hottocold} shows the reverse process, for which $f(Q_2,T_2) \leq f(Q_1,T_1)$, so
$f(Q_1,T_1)=f(Q_2,T_2)$ when ${Q_1 \over T_1}={Q_2 \over T_2}$.  Writing $\alpha={T_1 \over T_2}$  this leads to $f(Q,T)=f(\alpha Q, \alpha T)$.  As this must hold for all $T_1$ and $T_2$, and so for all $\alpha$
\begin{equation}\label{eq:tempfluc}
f(Q,T)=f\left({Q \over T}\right)
\end{equation}
The remaining four diagrams are essentially the same as the diagrams in Figures \ref{fg:kelvintoclausius}, \ref{fg:engtokelv} and \ref{fg:pumptokelv}, except they involve a Kelvin fluctuation from the higher temperature heat bath.  Comparison of these processes again leads to Equation \ref{eq:tempfluc}.

\subsection{Fluctuation Friendly Second Law}
Combining the result from Section \ref{ss:tempfluc}, with those from Sections \ref{ss:kelvtoclaus} to \ref{ss:pumpeng}, it is now possible to state the fluctuation compatible generalizations of the formulations of the Second Law of Thermodynamics given in Section \ref{s:phenom}
\begin{itemize}
\item \textbf{Kelvin}: There is no process, whose sole result is the extraction of a quantity of heat, $Q$, from a heat bath at temperature $T$, and its conversion to work, which can occur with probability $p$, unless:
\[
p \leq f \left( {Q \over T} \right)
\]
\item \textbf{Clausius}: There is no process, whose sole result is the extraction of a quantity of heat, $Q$, from a heat bath at temperature $T_1$, and its transfer to a heat bath at temperature $T_2 > T_1$, which can occur with probability $p$, unless:
\[
p \leq f \left( Q \left( {1 \over T_1 } - { 1 \over T_2} \right) \right)
\]
\item \textbf{Heat Engine}: There is no cyclic process, operating solely as a heat engine between heat baths at temperatures $T_2 > T_1$, which can extract a quantity of heat, $Q$, from the hotter heat bath, with efficiency $n_E$ exceeding that of a reliable, reversible heat engine, $n_{CE}$, with probability $p$, unless:
\[
p \leq f \left( {Q \over T_1} \left(n_E-n_{CE}\right) \right)
\]
\item \textbf{Heat Pump}: There is no cyclic process, operating solely as a heat pump between heat baths at temperatures $T_2 > T_1$, which can use a quantity of work, $W$, with efficiency $n_P$ exceeding that of a reliable, reversible heat engine, $n_{CP}$, with probability $p$, unless:
\[
p \leq f \left( {W \over T_1} \left({n_P \over n_{CP}} -1\right) \right)
\]
\end{itemize}
These four formulations are logically equivalent, in the same manner that the four formulations of the fluctuation-free second law given in Section \ref{s:phenom} are logically equivalent.
\subsection{Kelvin-Clausius inequality.}
These four formulations can be expressed in the same way.  Combining a single fluctuation with Carnot pumps and engines connecting heat baths at multiple temperatures reveals that there is a more general formulation of the fluctuation laws.  Just as all four of the normal phenomenological laws may be seen as special cases of the law:
\begin{quote}
There is no process, whose sole result is the extraction of quantities of heat, $Q_i$, from heat baths at temperatures $T_i$, converting the net heat extracted into work, unless:
\begin{equation}\label{eq:clausiusineq}
\sum_i {Q_i \over T_i} \leq 0
\end{equation}
\end{quote}
so all of the fluctuation laws are special cases of:
\begin{quote}
There is no process, whose sole result is the extraction of quantities of heat, $Q_i$, from heat baths at temperatures $T_i$, converting the net heat extracted into work, which can occur with probability $p$, unless:
\begin{equation}\label{eq:kelvinclausius}
p \leq f \left( \sum_i {Q_i \over T_i}\right)
\end{equation}
\end{quote}
The general formulation should make clear the role that Carnot cycles plays within the derivation of the specific fluctuation laws.  Carnot pumps and engines connecting a number of different heat baths are able to reversibly move heat between them in any combination provided the net effect is $\sum_i {Q_i \over T_i}=0$.  Any given fluctuation can therefore be converted into another fluctuation, involving different heat baths, but which has the same value of $\sum_i {Q_i \over T_i}$.
\subsection{Combining fluctuations}\label{ss:combfluc}
The next stage is to consider combining fluctuations, by diagrams involving more than one fluctuation.  As it turns out, only two diagrams, Figure \ref{fg:twoforone} are required to deduce the general relationship.
\doublepict{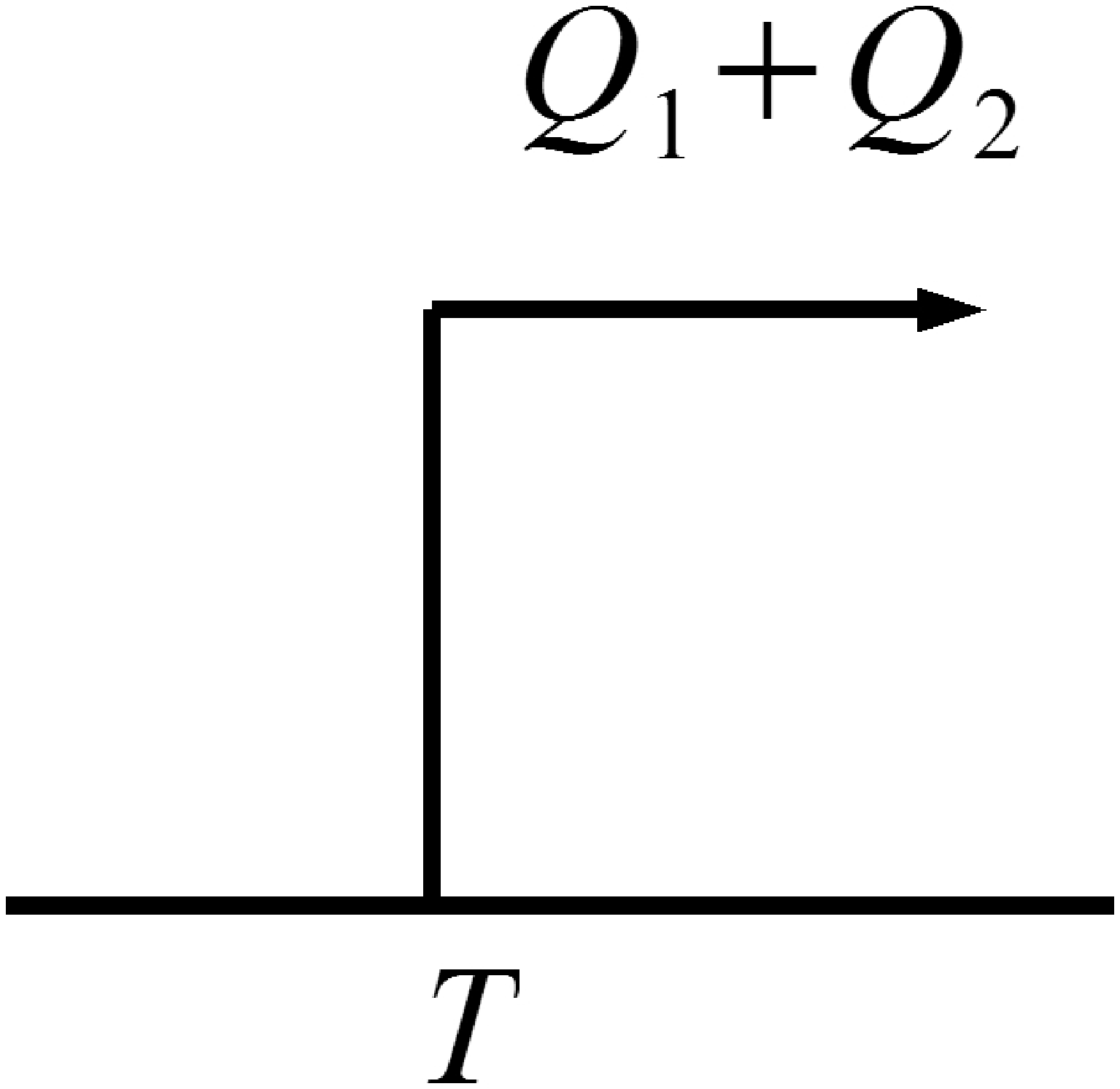}{kelvone}{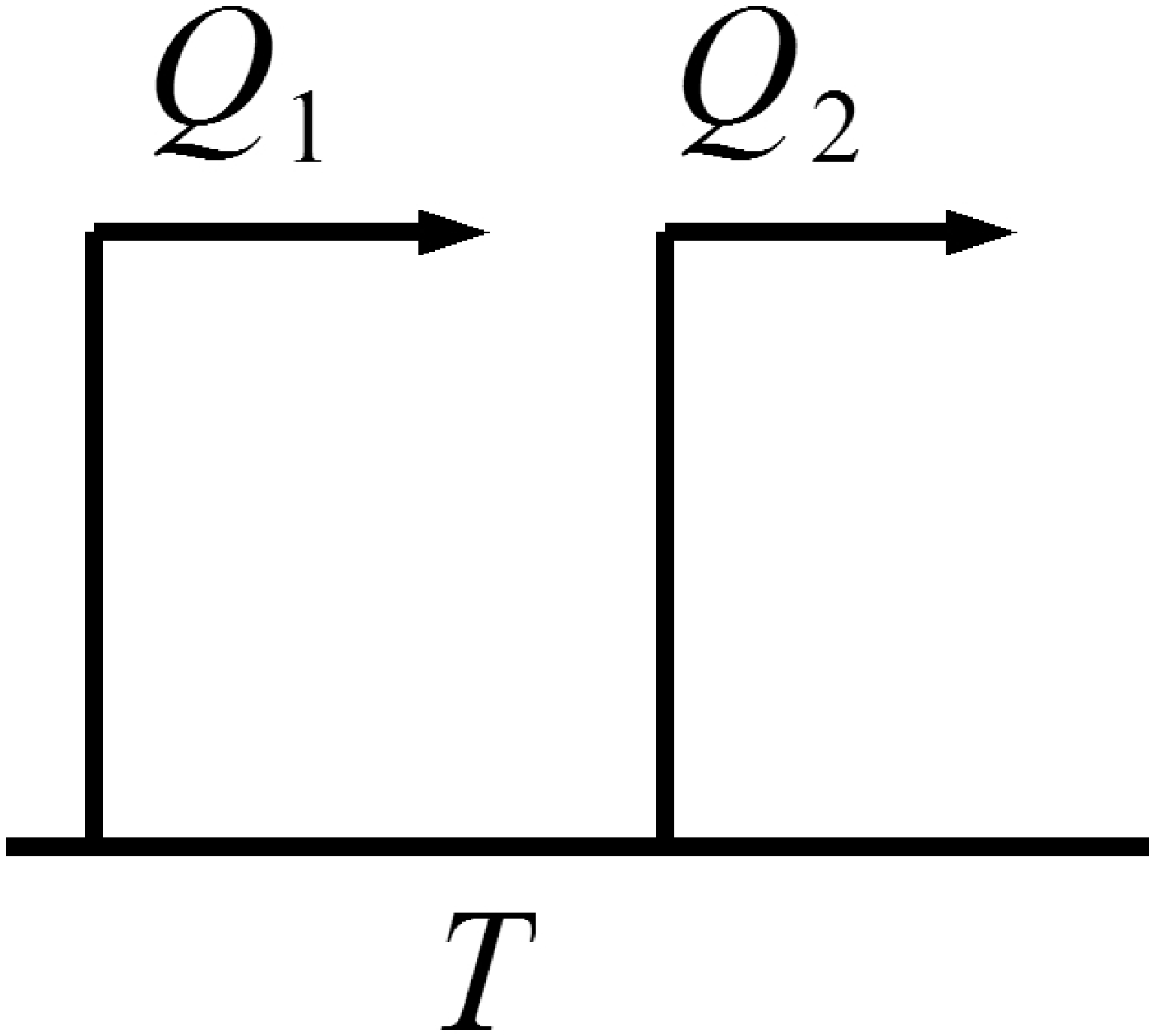}{kelvtwo}{Combining Kelvin Fluctuations}{twoforone}{0.2}

In Figure \ref{fg:kelvone} there is a single Kelvin fluctuation resulting in $Q_1+Q_2$ heat extracted from a heat bath at temperature $T$.  One possible way of this happening is if two independent processes occur, each from heat baths at temperature $T$, resulting in two separate Kelvin fluctuations, extracted $Q_1$ and $Q_2$ heat, respectively.  Figure \ref{fg:kelvtwo} gives a process by which $Q_1+Q_2$ can be extracted, so the minimal probability of a Kelvin fluctuation of that size cannot be less that the probability of the two independent fluctuations both occurring:
\begin{equation}
f\left({Q_1 \over T}+{Q_2 \over T}\right) \geq f\left({Q_1 \over T}\right)f\left({Q_2 \over T}\right)
\end{equation}
As this must happen for all $Q_1,Q_2,T$
the fluctuation law must satisfy the general functional inequality\footnote{This may be converted into a more familiar form using $F(x)=-\ln [ f(x) ]$ to get $F(x)+F(y) \geq F(x+y)$.  In passing, it may also be noted that if $f(x)$ is differentiable, then it can be shown from Equation \ref{eq:funcineq} that $f^\prime(x) \geq f(x)f^\prime(0)$ and $f^{\prime\prime}(0) \geq f^\prime(0)^2$.}
\begin{equation}\label{eq:funcineq}
f(x+y)\geq f(x)f(y)
\end{equation}
This leads directly to the general equation
\begin{equation}\label{eq:superexp}
f \left( \sum_i {Q_i \over T_i}\right) \geq \prod_i f\left({Q_i \over T_i}\right)
\end{equation}
that would also be deduced from considering diagrams with multiple fluctuations and with Carnot pumps and engines operating between multiple heat baths.

This property in itself can be used to demonstrate that, if there exists some $x=x_0 >0$ such that $f(x_0)=0$ then it must be the case that $\forall x>0, \; f(x)=0$, i.e. fluctuations must be possible at all scales, if they are possible on any scale.  Intuitively this should be obvious: provided a small fluctuation  can occur with a non-zero probability, $p$, then accumulating $n$ such fluctuations into a fluctuation $n$ times large is always possible with probability $p^n$.  Any size of fluctuation may occur with small, but non-zero probability, provided $n$ is large enough.

If it were the case that accumulating small fluctuations was the optimum process for obtaining a large fluctuations, then:
\begin{equation}
f \left( \sum_i {Q_i \over T_i}\right) = \prod_i f\left({Q_i \over T_i}\right)
\end{equation}
This requires $f(x+y)=f(x)f(y)$.  Provided $f$ is a continuous function, this has a unique solution:
\begin{equation}\label{eq:exponential}
f \left( \sum_i {Q_i \over T_i}\right) = e^{-\lambda \left( \sum_i {Q_i \over T_i}\right)}
\end{equation}
where $\lambda$ is a universal constant whose value would need determining experimentally to be the reciprocal of Boltzmann's constant: $\lambda=k^{-1}$.

It is, perhaps, surprising that such a familiar function within statistical mechanics might be obtained from the purely phenomenological arguments followed here!  Unfortunately, there seems no strong reason to demand that a large fluctuation cannot, in principle, be more probable than getting an equivalent sized fluctuation through the accumulation of a large number of small fluctuations.  It may, on the arguments considered so far, simply be the case that large fluctuations can spontaneously occur, with a higher probability.

Equation \ref{eq:exponential} is not the only possibility.  The restrictions on the form of $f(x)$ are
\begin{eqnarray}
f(x) & \geq 0 & \\ f(x) &=1& \forall x \leq 0 \\  \partxy{f}{x} & \leq 0 & \forall x >0 \\ f(x+y) &\geq f(x)f(y) & \forall x, y> 0
\end{eqnarray}
Other functions which could satisfy all these requirements include:
\begin{enumerate}
\item
\begin{equation}
f(x)= {1 \over {1+\sum_n a_n x^n}}
\end{equation}
will satisfy all the conditions specified whenever $n! a_n \leq m! l! a_m a_l$ for all $n=m+l$.  Specific cases include:
\begin{enumerate}
\item  $n! a_n = m! l! a_m a_l$.  This leads to $a_n= {{(a_1)^n} \over {n!}}$
\begin{equation}
f_e(x)=e^{-a_1 x}
\end{equation}
\item For all $n>1$, let $a_n=0$
\begin{equation}
f_i(x)={1 \over {1+a_1 x}}
\end{equation}
\item If some $f(x)$ that satisfies the conditions, then $g(x)=f^n(x)$, with $n>1$ will satisfy the conditions, so
\begin{equation}
f_q(x)={1 \over {(1+a_1 x)^{1/a_1}}}
\end{equation}
with $0 \leq a_1 \leq 1$.
\end{enumerate}
\item An even slower falling function such as
\begin{equation}
f_l(x)= {1 \over {1+ \ln (1+a x)}}
\end{equation}
can also satisfy the requirements.
\end{enumerate}
\section{Fluctuations and Entropy}\label{s:entropy}
In Section \ref{s:fluctuation} it was shown that the Kelvin-Clausius-Carnot versions of the second law, formulated in terms of cyclic processes and heat baths, can be generalised in a consistent way to include fluctuation phenomena.  However, phenomenological thermodynamics does not become genuinely powerful until Equation \ref{eq:clausiusineq} is used to define a non-decreasing, global function of state called entropy.  With fluctuations possible, it is clear that any such globally defined function of state can decrease with some probability.  In this Section it is shown that it is still possible to define a meaningful entropy function, with a relationship to the fluctuation law in Equation \ref{eq:kelvinclausius}.

\subsection{Phenomenological Entropy}
The Kelvin-Clausius inequality:
\begin{quote}
There is no process, whose sole result is the extraction of quantities of heat, $Q_i$, from heat baths at temperatures $T_i$, converting the net heat extracted into work, unless:
\[
\sum_i {Q_i \over T_i} \leq 0
\]
\end{quote}
immediately implies that, if there exists a process, whose sole result is to transform state $A$, into state $B$, while extracting quantities of heat, $Q^{(AB)}_i$, from heat baths at temperatures $T_i$, then there is no process whose sole result can be to transform state $B$ into state $A$, while extracting quantities of heat, $Q^{(BA)}_i$, from heat baths at temperatures $T_i$, unless:
\begin{equation}\label{eq:abcycle}
\sum_i {Q^{(AB)}_i \over T_i} + \sum_j {Q^{(BA)}_i \over T_i} \leq 0
\end{equation}
It is a straightforward mathematical construction (see Appendix \ref{appendix}) to show this implies the existence of a non-empty convex\footnote{For any two $S,S^\prime \in \{S_\theta(X)\}$ then for any $0 \leq p \leq 1$, it is the case that $pS+(1-p)S^\prime \in \{S_\theta(X)\}$.} set of functions of state, $\{S_\theta(X)\}$, which each satisfy the following condition:
\begin{quote}
If there exists a process, whose sole result is to transform state $A$ into state $B$, while extracting quantities of heat, $Q^{(AB)}_i$, from heat baths at temperatures $T_i$, then
\begin{equation}
S_\theta(A) \leq S_\theta(B) - \sum_i {Q^{(AB)}_i \over T_i}
\end{equation}
\end{quote}
The functions $S_\theta(X)$ will be referred to as thermodynamic entropies.

The expression of the phenomenological second law, in terms of these thermodynamic entropies, is
\begin{quote}
There exist functions of the thermodynamic state $\{S_\theta(X)\}$, such that for any two thermodynamic states $A$ and $B$, there is no process, whose sole result is to transform state $A$ into state $B$, while extracting quantities of heat $Q_i^{(AB)}$, from heat baths at temperatures $T_i$, unless
\begin{equation}
S_\theta(A) \leq S_\theta(B) - \sum_i {Q_i^{(AB)} \over T_i}
\end{equation}
for all $S_\theta(X)$.
\end{quote}
In an adiabatic process, no heat is extracted or generated in any heat bath, so this requires $S_\theta(A)
\leq S_\theta(B)$.

As this result must also hold for processes which transform $B$ into $A$, then
\begin{equation}
\sum_i {Q^{(AB)}_i \over T_i} \leq S_\theta(B)-S_\theta(A) \leq -\sum_i {Q^{(BA)}_i \over T_i}
\end{equation}
This must hold for all processes, so the set $\{S_\theta(X)\}$ is bounded by the processes which maximise the quantities $\sum_i \frac{Q^{(AB)}_i}{T_i}$ and $\sum_i \frac{Q^{(BA)}_i}{T_i}$.

If the two states $A$ and $B$ can be connected by a reversible cycle, then the maximum reached is
\begin{equation}
\sum_i {Q^{(AB)}_i \over T_i} + \sum_i {Q^{(BA)}_i \over T_i} =0
\end{equation}
in which case the entropy difference between the two states is fixed to be the same value for all functions in $\{S_\theta(X)\}$:
\begin{equation}
S_\theta(B)-S_\theta(A) =\sum_i \frac{Q^{(AB)}_i}{T_i}= -\sum_i \frac{Q^{(BA)}_i}{T_i}
\end{equation}
If all states can be connected by reversible cycles, then there is a single function, unique up to an additive constant.  It is important to note that reversibility is required for the uniqueness of the entropy function, but is not necessary to prove the existence of a non-decreasing set of entropy functions.

\subsection{Fluctuation Entropy law}
The existence of the fluctuation law does not prevent the derivation of the existence of the thermodynamic entropy functions $\{S_\theta(X)\}$.  Their significance is restricted to reliable (i.e. probability one) processes.  Unfortunately it does not immediately follow that a fluctuation law can be deduced constraining the probability of a reduction in thermodynamic entropy.

An essential stage in the deduction of a law relating entropy to fluctuations, is the identification of an appropriate inequality for closed cycles incorporating any two states, such as Equation \ref{eq:abcycle}, but for cycles involving fluctuations.  Such an inequality cannot be directly obtained from the fluctuation law.

The fluctuation law, Equation (\ref{eq:kelvinclausius}), implies that, if there exists a process, whose sole result is to transform state $A$, into state $B$, while extracting quantities of heat, $Q^{(AB)}_i$, from heat baths at temperatures $T_i$, and which can occur with probability $p_{AB}$, then there is no process whose sole result can be to transform state $B$ into state $A$, while extracting quantities of heat, $Q^{(BA)}_i$, from heat baths at temperature $T_i$, which can occur with probability $p_{BA}$, unless:
\begin{equation}
p_{AB}p_{BA} \leq  f\left( \sum_i \frac{Q^{(AB)}_i}{T_i} + \sum_i \frac{Q^{(BA)}_i }{T_i} \right)
\end{equation}
Inverting the function gives:
\begin{equation}\label{eq:pabpba}
f^{-1}(p_{AB}p_{BA}) \geq \sum_i \frac{Q^{(AB)}_i }{T_i} + \sum_i \frac{Q^{(BA)}_i}{T_i}
\end{equation}
However, the relationship $f(x+y) \geq f(x)f(y)$, when inverted, yields
\begin{equation}
f^{-1}(pq) \geq f^{-1}(p)+f^{-1}(q)
\end{equation}
and this does not allow the deduction of a suitable inequality.

\subsubsection{Reliable Paths}
To proceed further, it is necessary to consider reliable paths between $A$ and $B$.  Let $q^{(AB)}_i$ be the heat generated in heat baths at temperatures $T_i$, for a process that can occur with probability one, and whose sole effect, apart from extracting heat from heat baths and converting them to work, is to transform state $A$ into state $B$.  It follows that there is no process, whose sole result is to transform state $B$ into state $A$, while extracting quantities of heat $Q^{(BA)}_i$ from heat baths at temperatures $T_i$, which can occur with probability $p_{BA}$, unless
\begin{equation}\label{eq:optflucab}
f^{-1}(p_{BA}) \geq \sum_i {q^{(AB)}_i \over T_i} + \sum_i {Q^{(BA)}_i \over T_i}
\end{equation}
Similarly, if $q^{(BA)}_i$ is the heat generated in heat baths at temperatures $T_i$, for a process that can occur with probability one, whose sole effect, apart from extracting heat from heat baths and converting them to work, is to transform state $B$ into state $A$, then there is no process, whose sole result is to transform state $A$ into state $B$, while extracting quantities of heat $Q^{(AB)}_i$ from heat baths at temperatures $T_i$, which can occur with probability $p_{AB}$, unless
\begin{equation}\label{eq:optflucba}
f^{-1}(p_{AB}) \geq \sum_i {Q^{(AB)}_i \over T_i} + \sum_i {q^{(BA)}_i \over T_i}
\end{equation}
It is immediately possible to deduce both that
\begin{equation}\label{eq:minpathcycle}
\sum_i {q^{(AB)}_i \over T_i}+ \sum_i {q^{(BA)}_i \over T_i} \leq 0
\end{equation}
(by using a process for which either $p_{AB}=1$ or $p_{BA}=1$) and that
\begin{equation}\label{eq:minpathfluc}
f^{-1}(p_{AB})+f^{-1}(p_{BA})  \geq \sum_i \frac{Q^{(AB)}_i + q^{(BA)}_i + q^{(AB)}_i + Q^{(BA)}_i }{T_i}
\end{equation}
Equation \ref{eq:minpathcycle} implies the existence of the thermodynamic entropies $\{S_\theta(X)\}$, as before. Equation \ref{eq:minpathfluc} implies the existence of a convex set of functions of state $\{S_\phi(X)\}$, which will be called the fluctuation entropies, and which all satisfy
\begin{widetext}
\begin{equation}
\sum_i {{Q^{(AB)}_i+q^{(AB)}_i }\over T_i} -f^{-1}(p_{AB}) \leq S_\phi(B)-S_\phi(A) \leq f^{-1}(p_{BA}) -\sum_i {{Q^{(BA)}_i +q^{(BA)}_i}\over T_i}
\end{equation}
\end{widetext}
In order to narrow down the range of permissible entropies, the terms $\sum_i {q^{(AB)}_i \over T_i}$ and $\sum_i {q^{(BA)}_i \over T_i}$ should each be as large as possible, subject to the constraint of Equation \ref{eq:minpathcycle}.

This produces the following entropy fluctuation law:
\begin{quote}
Let $\mathcal{Q}^{(AB)}_i$ be the heats extracted from heat bath at temperatures $T_i$, by a process, which occurs with probability one, whose sole other result is to transform state $A$ into state $B$, and which maximises the value of $\sum_i \frac{\mathcal{Q}^{(AB)}_i }{T_i}$ over all such processes.

There exists single valued functions of state $\{S_\phi(X)\}$, such that, if there exists a process occurring with probability $p$, whose sole result is to transform state $A$ into state $B$, while extracting quantities of heat, $Q^{(AB)}_i$, from heat baths at temperatures $T_i$, then
\begin{equation}
S_\phi(A) \leq S_\phi(B) +f^{-1}(p) - \sum_i {\mathcal{Q}^{(AB)}_i \over T_i} - \sum_i {Q^{(AB)}_i \over T_i}
\end{equation}
\end{quote}

To restrict these to a unique function $S_\phi(X)$ requires that there exist cycles\footnote{When dealing with fluctuations, a cycle is a process for which the system starts in state $A$ with certainty, reaches the state $B$ with probability $p_{AB}$, and then the conditional probability for returning to state $A$, given that it reached state $B$, is $P_{BA}$.} for which
\begin{equation}\label{eq:revfluccycle}
 f^{-1}(p_{AB}) +f^{-1}(p_{BA}) =\sum_i \frac{Q^{(AB)}_i + \mathcal{Q}^{(AB)}_i + Q^{(BA)}_i +\mathcal{Q}^{(BA)}_i }{T_i}
\end{equation}

\subsubsection{Reversible Paths}
If it is the case that the equality in Equation \ref{eq:minpathcycle} is met, then Equation \ref{eq:minpathfluc} takes the form
\begin{equation}\label{eq:minpathfluc2}
f^{-1}(p_{AB})+f^{-1}(p_{BA})  \geq \sum_i {Q^{(AB)}_i \over T_i}+ \sum_i {Q^{(BA)}_i \over T_i}
\end{equation}
and
\begin{equation}
\sum_i {Q^{(AB)}_i \over T_i} -f^{-1}(p_{AB}) \leq S_\phi(B)-S_\phi(A) \leq f^{-1}(p_{BA}) -\sum_i {Q^{(BA)}_i \over T_i}
\end{equation}
This is not sufficient to ensure $S_\phi(B)-S_\phi(A)$ is unique.  However, in this case, it is possible to deduce the existence of the globally unique thermodynamic entropy from the reliable paths
\begin{equation}
S_\theta(B)-S_\theta(A) =\sum_i {\mathcal{Q}^{(AB)}_i \over T_i}= -\sum_j {\mathcal{Q}^{(BA)}_i \over T_i}
\end{equation}
for which Equations \ref{eq:optflucab} and \ref{eq:optflucba} give
\begin{equation}\label{eq:thermentfluc}
\sum_i {Q^{(AB)}_i \over T_i} -f^{-1}(p_{AB}) \leq  S_\theta(B)-S_\theta(A) \leq f^{-1}(p_{BA})- \sum_i {Q^{(BA)}_i \over T_i}
\end{equation}
and $S_\theta(X) \in \{S_\phi(X)\}$.  If both $S_\theta(X)$ and $S_\phi(X)$ are uniquely defined, then $\sum_i {Q^{(AB)}_i \over T_i} -f^{-1}(p_{AB}) +  \sum_i {Q^{(BA)}_i \over T_i} -f^{-1}(p_{BA})=0$, in which case $S_\theta(X)=S_\phi(X)$.    However, in general, if the thermodynamic entropies $\{S_\theta(X)\}$ are not restricted to a single globally unique function, then there may exist $S_\theta(X) \notin \{S_\phi(X)\}$.

It is worth noting that Equation \ref{eq:thermentfluc} implies
\begin{equation}
\sum_i {Q^{(AB)}_i \over T_i} -f^{-1}(p_{AB}) +  \sum_i {Q^{(BA)}_i \over T_i} - f^{-1}(p_{BA}) \leq	 0
\end{equation}
If, on the other hand, Equation \ref{eq:revfluccycle} holds for fluctuation cycle, for which also $\sum_i {q^{(AB)}_i \over T_i}+ \sum_i {q^{(BA)}_i \over T_i}  < 0$, then this requires
\begin{equation}\label{eq:badentfluc}
\sum_i {Q^{(AB)}_i \over T_i} -f^{-1}(p_{AB}) +  \sum_i {Q^{(BA)}_i \over T_i} - f^{-1}(p_{BA}) > 0
\end{equation}
In other words, if there exist \textit{any} fluctuations from state $A$ to state $B$, and vice versa, that can define a unique fluctuation entropy difference $S_\phi(B)-S_\phi(A)$ when combined with a reliable but \textit{irreversible} cyclic path between $A$ and $B$, then it must be the case that there are \textit{no} reliable, reversible cyclic paths between states $A$ and $B$.  The existence of a globally unique $S_\phi(X)$ that is not simultaneously a globally unique $S_\theta(X)$ would imply reliable, reversible processes cannot exist.

Reliable, reversible cycles imply an entropy fluctuation law:
\begin{quote}
There exists single valued functions of state $\{S_\phi(X)\}$, such that, if there exists a cyclic process, occurring with probability one, operating between states $A$ and states $B$, with a zero net extraction of heat over the cycle, then for any other process, occurring with probability $p$, whose sole result is to transform state $A$ into state $B$, while extracting quantities of heat, $Q^{(AB)}_i$, from heat baths at temperatures $T_i$, then
\begin{equation}
S_\phi(A) \leq S_\phi(B) +f^{-1}(p)- \sum_i {Q^{(AB)}_i \over T_i}
\end{equation}
and there is a globally unique thermodynamic entropy $S_\theta(X) \in \{S_\phi(X)\}$
\end{quote}

\subsubsection{Exponential Statistics}
Finally, note that if the fluctuation law takes the exponential form discussed in Section \ref{ss:combfluc}, then
\begin{equation}
f^{-1}(pq) = f^{-1}(p)+f^{-1}(q)
\end{equation}
so Equation \ref{eq:pabpba} leads immediately to
\begin{equation}\label{eq:pabpluspba}
f^{-1}(p_{AB})+f^{-1}(p_{BA}) \geq \sum_i {Q^{(AB)}_i \over T_i} + \sum_j {Q^{(BA)}_j \over T_j}
\end{equation}
This gives  Equation \ref{eq:thermentfluc} without needing the existence of reliable paths.  This implies there exists a convex set of fluctuation entropies $\{S_\eta(X)\} \subseteq \{S_\phi(X)\}$  satisfying
\begin{equation}
\sum_i {Q^{(AB)}_i \over T_i} -f^{-1}(p_{AB}) \leq S_\eta(B)-S_\eta(A) \leq f^{-1}(p_{BA}) -\sum_j {Q^{(BA)}_j \over T_j}
\end{equation}
Uniquely defining an $S_\eta(X)$ entropy would require $\sum_i {Q^{(AB)}_i \over T_i} -f^{-1}(p_{AB}) + \sum_j {Q^{(BA)}_j \over T_j} -f^{-1}(p_{BA}) =0$, but this does not necessarily uniquely define either $S_\phi(X)$ or $S_\theta(X)$.  In this case, however, if a unique $S_\phi(X)$ does exist then it is necessarily equal to a unique $S_\theta(X)$, and vice versa.
\section{From Fluctuations to Statistical Mechanics}\label{s:statmech}
The possible relationship of the fluctuation spectrum $f$ to statistical mechanics will now be briefly explored.  It will be assumed throughout this Section that a globally unique entropy $S(X)=S_\theta(X)=S_\phi(X)$ can be determined, and only a single heat bath at temperature $T$ will be used.  The entropy fluctuation law now takes the form:
\begin{quote}
There exists a single valued function of state $S(X)$, such that for any process, occurring with probability $p$, whose sole result is to transform state $A$ into state $B$, while extracting quantities of heat, $Q^{(AB)}$, from heat baths at temperatures $T$, then
\begin{equation}
S(A) \leq S(B) +f^{-1}(p)-  {Q^{(AB)} \over T}
\end{equation}
\end{quote}

Suppose the system is in an initial state, with entropy $S$, internal energy $E$, and is subject to a process during which it fluctuates to state $\alpha$ with probability $p_\alpha$.  During the course of the process, heats $Q_{\alpha}$ are generated in heat baths at temperatures $T$ and requires work $W_\alpha$ to be performed.

By conservation of energy, the internal energy of state $\alpha$ is
\begin{equation}
E_\alpha=E+W_\alpha -Q_\alpha
\end{equation}
By the entropy fluctuation law, the entropy of state $\alpha$ must obey
\begin{equation}\label{eq:statentineq}
S \leq S_\alpha +f^{-1}(p_\alpha)- {Q_\alpha \over T}
\end{equation}
This equation must hold for each possible fluctuation away from the initial state, so that
\begin{equation}\label{eq:statentsumineq}
S \leq \sum_\alpha p_\alpha \left(S_\alpha +f^{-1}(p_\alpha)- {Q_\alpha \over T}\right)
\end{equation}
necessarily holds.  The form of this constraint is very suggestive of entropy functions that occur in statistical mechanics.

\subsection{Maximal Fluctuations}
The definition of the $f$ function is such that there must exist some process for which the equality in
Equation \ref{eq:statentineq} is met:
\begin{eqnarray}\label{eq:statenteq}
S & = & S_\alpha +f^{-1}(p_\alpha)- {Q_\alpha \over T} \\
p_\alpha &=& f \left( \left(S- {E \over T} \right) -\left(S_\alpha - {{E_\alpha-W_\alpha} \over T}\right) \right)
\end{eqnarray}

However, there is no guarantee that a single process can exist which achieves the maximum fluctuation for every possible outcome.  If such a process did exist, then
\begin{equation}
S = \sum_\alpha p_\alpha \left(S_\alpha +f^{-1}(p_\alpha)-  {Q_{\alpha} \over T}\right)
\end{equation}
would hold.

This similarity to statistical mechanics is brought even closer under two conditions:
\begin{enumerate}
\item If a set of maximal fluctuations occur which do not generate heat, on average, then $\sum_\alpha p_\alpha {Q_\alpha \over T}=0$.  The entropy formula then becomes:
\begin{equation}
S = \sum_\alpha p_\alpha \left(S_\alpha +f^{-1}(p_\alpha)\right)
\end{equation}
\item If a set of maximal fluctuations can take place, without requiring external work to be performed $(W_\alpha=0)$ then:
\begin{equation}
p_\alpha=f\left( \left(S- {E \over T} \right) -\left(S_\alpha - {E_\alpha \over T}\right) \right)
\end{equation}
or $p_\alpha=f({ {F - F_\alpha} \over T})$, where
\begin{eqnarray}
F&=&TS- E\\
F_\alpha &=&TS_\alpha - E_\alpha
\end{eqnarray}
\end{enumerate}
\subsection{Example fluctuation laws}
Let us consider the functions from Section \ref{ss:combfluc}

\begin{enumerate}
\item $f(x)= {1 \over {1+\sum_n a_n x^n}}$
\begin{enumerate}
\item $f_e(x)=e^{-a_1 x}$.  This generates the familiar Gibbs canonical statistics.
\begin{eqnarray}
f^{-1}_e(p)&=& -{1 \over a_1} \ln p \\
S &=& \sum_\alpha p_\alpha S_\alpha -{1 \over a_1} p_\alpha \ln p_\alpha  \\
p_\alpha &=& {1 \over Z_e} e^{-a_1 F_\alpha / T }
\end{eqnarray}
with $Z_e=e^{a_1 F / T }=\sum_\alpha e^{-a_1 F_\alpha / T }$
\item $f_i(x)= (1+a_1 x)^{-1}$
\begin{eqnarray}
f^{-1}_i(p)&=&{1 \over a_1} \left(p^{-1}-1\right)\\
S &=& \sum_\alpha p_\alpha S_\alpha -{1 \over a_1} (N-1)  \\
p_\alpha &=& {1 \over Z_i} (1+a_1 \beta F_\alpha )^{-1}
\end{eqnarray}
with $N$ the number of distinct states in the summation, $Z_i=(1+a_1 F)$ and $\beta=1/(TZ_i)$.
\item $f_q(x)=(1+a_1 x)^{-1/a_1}$.  This generates statistics similar to the Tsallis non-extensive entropies.
\begin{eqnarray}
f^{-1}_q(p)&=&{1 \over a_1} \left(p^{-a_1}-1\right) \\
S &=& \sum_\alpha p_\alpha S_\alpha -{1 \over a_1} \left(1-\sum_\alpha p_\alpha^{1-a_1}\right)  \\
p_\alpha &=& {1 \over Z_q} (1+a_1 \beta F_\alpha )^{-1/a_1}
\end{eqnarray}
with $Z_q=(1+a_1 F/T)^{1/a_1}$ and $\beta=1/(TZ_q^{a_1})$.
\end{enumerate}
\item The slowly falling function $f_l(x)=(1+\ln (1+ax))^{-1}$ yields
\begin{eqnarray}
f_l^{-1}(p)&=&{1 \over a_1} \left( e^{(p^{-1} -1)}-1 \right) \\
S &=& \sum_\alpha p_\alpha S_\alpha +{1 \over a_1} \left(\sum_\alpha p_\alpha e^{(p^{-1} -1)}\right) - {1 \over a_1}  \\
p_\alpha &=& {1 \over Z_l} (1+\ln (1+a_1 \beta F_\alpha )^{1/Z_l})^{-1}
\end{eqnarray}
with $Z_l=1+\ln (1+a_1 F/T)$ and $\beta=e^{Z_l-1}/T$
\end{enumerate}
\section{Conclusion}\label{s:conclusion}
Starting from the physical intuition that larger thermal fluctuations must be less probable than smaller fluctuations, we have suggested a fluctuation law that states that for any given size of fluctuation, there is a non-trivial maximum probability of it occurring.  This simple suggestion proves surprisingly fruitful.  The equivalence of the Kelvin, Clausius and Carnot formulations of the phenomenological second law of thermodynamics is shown to naturally generalise to the fluctuation law, and further constrain it to be of the form:
\begin{quote}
There is no process, whose sole result is the extraction of quantities of heat, $Q_i$, from heat baths at temperatures $T_i$, converting the net heat extracted into work, which can occur with probability $p$, unless:
\begin{equation}
p \leq f \left( \sum_i {Q_i \over T_i}\right)
\end{equation}
\end{quote}
with the function $f$ further constrained by the requirement
\begin{equation}
f \left( \sum_i {Q_i \over T_i}\right) \geq \prod_i f\left({Q_i \over T_i}\right)
\end{equation}
If the underlying dynamics is found to be such that larger fluctuations can only occur through the accumulation of smaller fluctuations, then this requires the function to have the exponential form:
\begin{equation}
f \left( \sum_i {Q_i \over T_i}\right) = e^{-\lambda \left( \sum_i {Q_i \over T_i}\right)}
\end{equation}
It is interesting to note that the phenomenologically motivated approaches of Szilard and of Tisza and Quay\cite{Szi1925,TQ1963} to statistical mechanics derive the canonical distribution by making a similar assumption (see also \cite{Maroney2007a}).

We have further shown that the deduction of the existence of a non-decreasing thermodynamic entropy function of state may still be followed, to derive a fluctuation entropy function of state.  Under a similar kind of circumstance for which the thermodynamic entropy can be deduced to be globally unique, then the fluctuation entropy can be deduced to be globally unique.  Furthermore, if the thermodynamic and fluctuation entropies are both globally unique, then they are necessarily identical (up to an additive constant).  This holds out hope that more rigorously axiomatic developments of the thermodynamic entropy, such as that of Lieb and Yngvason\cite{LY1999}, may be generalized in a similar manner to incorporate fluctuation phenomena.

Some possible forms of the entropy fluctuation law have been investigated.  The exponential form naturally produces the Gibbs canonical distribution for thermal fluctuations.  Non-extensive entropies, such as the Tsallis entropy, can also be seen to arise naturally in this approach.  Further investigation  is needed to explore the consistency of different $f$ functions.  In particular, the requirement that the \textit{mean} heat extracted over a cycle is non-positive, $ \left\langle {\sum_i \frac{Q_i}{T_i}} \right\rangle \leq 0 $, may be expected to further constrain which functions are admissible.
\appendix*
\section{Entropy Functions for Irreversible Cycles}\label{appendix}
Suppose there exists a path dependant quantity, $\Omega^\lambda_{AB}$ (a property of a particular path $\lambda$, in a state space, from state $A$ to state $B$) well defined for all paths $\lambda$, states $A$ and states $B$, for which:
\begin{equation}
\forall \lambda, \lambda^\prime \;\; \Omega^\lambda_{AB}+\Omega^{\lambda^\prime}_{BA} \geq 0 \label{appclaus}
\end{equation}
and that there exists at least one path from each $A$ to each $B$ for which the corresponding value of $\Omega$ is finite, so that $\inf_\lambda \left[\Omega^\lambda_{AB}\right] < \infty$.  Then there exists a non-empty convex set of functions of state $\{S(X)\}$, such that for all paths $\lambda$ and states $A$ and $B$:
\begin{equation}
S(A) \leq S(B) +\Omega^\lambda_{AB} \label{appent}
\end{equation}

\textbf{Proof}:
Define $\Omega_{AB}=\inf_\lambda \left[\Omega^\lambda_{AB}\right]$.  So $\Omega^\lambda_{AB} \geq \Omega_{AB}$.

As $\Omega_{BA} < \infty$ and $\Omega_{AB} \geq -\Omega_{BA}$, then $\Omega_{AB} > -\infty$.

By definition, the minimum value of $\Omega$ going from $A$ to $C$ cannot be more than the value going from $A$ to $C$ via a path including $B$:
\begin{equation}
\Omega_{AC} \leq \Omega_{AB}+\Omega_{BC}
\end{equation}
so
\begin{eqnarray}
\Omega_{AC}-\Omega_{AB} & \leq & \Omega_{BC} \\
\Omega_{AB}-\Omega_{AC} & \geq & -\Omega_{BC}\\
\Omega_{AC}-\Omega_{BC} & \leq & \Omega_{AB} \\
\Omega_{BC}-\Omega_{AC} & \geq & -\Omega_{AB}
\end{eqnarray}

Define the set of functions of state $\{S_{iY}(X)\}$ by
\begin{eqnarray}
S_{+A}(X) &=& \Omega_{XA}\\
S_{-A}(X) &=& -\Omega_{AX}
\end{eqnarray}
These are clearly well defined, finite functions of state, and they exist, so the set $\{S_{iY}(X)\}$ is not empty.
Note that as $\Omega_{XX}=0$:
\begin{eqnarray}
\Omega_{XY} &=& S_{+Y}(X)-S_{+Y}(Y) \\
 &=&S_{-X}(X)-S_{-X}(Y)
 \end{eqnarray}
and
\begin{eqnarray}
S_{+A}(X) -S_{+A}(Y) &=& \Omega_{AX}-\Omega_{AY} \\
S_{-A}(X) -S_{-A}(Y) &=& -\Omega_{XA}+\Omega_{YA}
\end{eqnarray}

It follows that for any $A$,
\begin{eqnarray}
S_{+A}(X) -S_{+A}(Y) & \leq& \Omega_{XY} \leq \Omega^\lambda_{XY}\\
& \geq &-\Omega_{YX} \geq -\Omega^\lambda_{YX}\\
S_{-A}(X) -S_{-A}(Y) &\leq& \Omega_{XY} \leq \Omega^\lambda_{XY}\\
&\geq &-\Omega_{YX} \geq -\Omega^\lambda_{YX}
\end{eqnarray}
and it is then easily demonstrated that for any distribution $\sum_{iY} w(iY)=1$, $w(iY)\geq 0$, that the weighted function of state
\begin{equation}
S(X)=\sum_{iY}w(iY)S_{iY}(X)
\end{equation}
satisfies
\begin{equation}
S(A)-S(B) \leq \Omega^\lambda_{AB}
\end{equation}
as
\begin{widetext}
\begin{equation}
-\Omega^\lambda_{YX}
 \leq -\Omega_{YX}=S_{+X}(X)-S_{+X}(Y) \leq S_{iA}(X)-S_{iA}(Y) \leq S_{+Y}(X)-S_{+Y}(Y)=\Omega_{XY} \leq \Omega^{\lambda^\prime}_{XY}
 \label{appbigeq}
\end{equation}
\end{widetext}
Note, that the set $\{\sum_{iY}w(iY)S_{iY}(X)\}$ does not necessarily include all the functions which satisfy the inequality of Eq. ~(\ref{appent}).  It only demonstrates the existence of a non-empty set of such functions.

It is now a trivial matter to show from Equation \ref{appbigeq} that, whenever the equality in Equation \ref{appclaus} can be reached, that all functions in the set $\{\sum_{iY}w(iY)S_{iY}(X)\}$  (indeed, all functions satisfying Equation \ref{appent}) will give the same entropy difference between states $A$ and $B$.  By extension, if the equality in Equation \ref{appclaus} can be reached for all pairs of states, then there is a single function, $S(X)$, unique up to an additive constant.
\begin{acknowledgments}
I would like to thank Harvey Brown, John Norton, Tony Short, Jos Uffink and Steve Weinstein for discussions and suggestions that have influenced the development of this paper, and an anonymous referee for helpful comments.  Research at the Perimeter Institute for Theoretical Physics is supported in part by the Government of Canada through NSERC and by the Province of Ontario through MRI.
\end{acknowledgments}

\end{document}